\documentclass[12pt]{article}
\pdfoutput=1
\usepackage{times}
\usepackage{epsfig}
\usepackage{amsmath}
\usepackage{amsfonts}
\usepackage{color}

\def\beq{\begin{equation}}
\def\eeq{\end{equation}}
\def\bea{\begin{eqnarray}}
\def\eea{\end{eqnarray}}
\def\beqn{\begin{eqnarray}} \def\eeqn{\end{eqnarray}}
\def\beeq{\begin{eqnarray}}
\def\eeeq{\end{eqnarray}}

\def\ep{\epsilon}

\def\nn{\nonumber}
\def\Eq#1{Eq.~(\ref{#1})}
\def\ln#1{\mathrm{log}\left(#1\right)}

\def\li#1{\mathrm{Li_2}\left(#1\right)}
\def\bp{p\hspace{-.42em}/\hspace{-.07em}}
\def\bq{q\hspace{-.42em}/\hspace{-.07em}}

\def\td#1{\tilde{\delta}\left(#1\right)}

\newcommand{\la}{\langle}
\newcommand{\ra}{\rangle}

\def\qb{\mathbf{q}}
\def\pb{\mathbf{p}}

\def\M#1{{\cal M}^{(#1)}} 
\def\MD#1{{\cal M}^{(#1) \, \dagger}}

\def\qq{{q \bar q}}
\def\qqg{{q \bar q g}}

\newcommand\as{\alpha_{\mathrm{S}}}

\newcommand\muUV{\mu_{\rm UV}}

\def\uv{{\rm UV}}
\def\ir{{\rm IR}}
\def\cnt{{\rm cnt}}
\def\cg{c_{\Gamma}}
\def\cgt{\widetilde{c}_{\Gamma}}
\def\gs{g_{\rm S}}
\def\as{\alpha_{\rm S}}
\def\aas{\frac{\as}{4\pi}}
\def\v{{\rm V}}
\def\r{{\rm R}}
\def\f{{\rm f}}
\def\b{{\rm b}}
\def\re{{\rm Re}}
\def\im{{\rm Im}}
\def\xiu{\xi_{1,0}}
\def\xid{\xi_{2,0}}
\def\xit{\xi_{3,0}}
\def\xiuv{\xi_{\uv}}
\def\SMS{S_{\ep}^{\overline {\rm MS}}}
\def\Se{{\widetilde S}_{\ep}}
\def\nlo{{\rm NLO}}

\begin{document}

\begin{titlepage}
\renewcommand{\thefootnote}{\fnsymbol{footnote}}
\begin{flushright}
     IFIC/15-73
     \end{flushright}
\par \vspace{10mm}
\begin{center}
{\LARGE \bf
Four-dimensional unsubtraction \vspace{2mm} \\ from the loop-tree duality}
\end{center}
\par \vspace{2mm}
\begin{center}
{\bf 
Germ\'an F. R. Sborlini~$^{(a,b)}$\footnote{E-mail: gfsborlini@df.uba.ar},
F\'elix Driencourt-Mangin~$^{(a)}$\footnote{E-mail: felix.dm@ific.uv.es},
Roger J. Hern\'andez-Pinto~$^{(a,c)}$\footnote{E-mail: rogerjose.hernandez@ific.uv.es}}
and {\bf
Germ\'an Rodrigo~$^{(a)}$\footnote{E-mail: german.rodrigo@csic.es}}
\vspace{5mm}

${}^{(a)}$ Instituto de F\'{\i}sica Corpuscular, 
Universitat de Val\`{e}ncia -- 
Consejo Superior de Investigaciones Cient\'{\i}ficas, 
Parc Cient\'{\i}fic, E-46980 Paterna, Valencia, Spain \\
${}^{(b)}$Departamento de F\'\i sica and IFIBA, FCEyN, Universidad de Buenos Aires, \\
(1428) Pabell\'on 1 Ciudad Universitaria, Capital Federal, Argentina \\
${}^{(c)}$ Facultad de Ciencias F\'isico-Matem\'aticas, Universidad Aut\'onoma de Sinaloa, \\
Ciudad Universitaria, CP 80000, Culiac\'an, Sinaloa, M\'exico \\
\end{center}

\par \vspace{2mm}
\begin{center} {\large \bf Abstract} \end{center}
\begin{quote}
We present a new algorithm to construct a purely four dimensional representation of 
higher-order perturbative corrections to physical cross-sections at next-to-leading order (NLO). 
The algorithm is based on the loop-tree duality (LTD), and it is implemented by introducing a suitable mapping 
between the external and loop momenta of the virtual scattering amplitudes, and the external momenta of the
real emission corrections. In this way, the sum over degenerate infrared states is
performed at integrand level and the cancellation of infrared divergences occurs locally
without introducing subtraction counter-terms to deal with soft and final-state collinear singularities.
The dual representation of ultraviolet counter-terms is also discussed in detail, in particular 
for self-energy contributions.  The method is first illustrated with the scalar three-point function, 
before proceeding with the calculation of the physical cross-section for $\gamma^* \to q \bar{q}(g)$, 
and its generalisation to multi-leg processes.  The extension to next-to-next-to-leading order (NNLO) is briefly commented. 

\end{quote}

\par \vspace{5mm}

\vspace*{\fill}

\end{titlepage}

\setcounter{footnote}{0}
\renewcommand{\thefootnote}{\fnsymbol{footnote}}

\section{Introduction}
\label{sec:intro}
Computation of higher-order perturbative corrections in quantum field theories is a very active research field in these days
in view of the high-quality data provided by the LHC. 
The standard approach to perform these calculations in perturbative QCD relies in the application of the subtraction formalism. 
According to Kinoshita-Lee-Nauenberg theorem~\cite{Kinoshita:1962ur, Lee:1964is} theoretical predictions in theories with massless particles 
can only be obtained after the definition of infrared-safe physical observables. These involve performing a sum over all degenerate states, 
which means adding together real and virtual contributions. After ultraviolet (UV) renormalisation of virtual scattering amplitudes, 
the remaining contributions develop infrared (IR) singularities that cancel when putting all the terms together. 
This implies that the IR divergent structure of real and virtual corrections are closely related: thus the subtraction method 
exploits the factorisation properties of QCD to define suitable subtraction IR counter-terms which mimic their singular behaviour. 

There are several variants of the subtraction method at NLO and beyond~\cite{Kunszt:1992tn,Frixione:1995ms, Catani:1996jh, Catani:1996vz, GehrmannDeRidder:2005cm,Catani:2007vq,Czakon:2010td,Bolzoni:2010bt,DelDuca:2015zqa,Boughezal:2015dva,Gaunt:2015pea}, 
which involve treating separately real and virtual contributions. However, from a computational point of view, these methods might not be efficient enough for multi-particle processes. The main reason for that is related to the fact that the 
final-state phase-space (PS) of the different contributions involves different numbers of particles. For instance, at NLO, 
virtual corrections with Born kinematics have to be combined with real contributions involving an additional final-state particle.
The IR counter-terms have to be local in the real PS, and analytically integrable over the extra-radiation 
factorised PS to properly cancel the divergent structure present in the virtual corrections. Building these counter-terms represents 
a challenge and introduces a potential bottleneck to efficiently carry out the IR subtraction for multi-leg multi-loop processes.

With the aim of eluding the introduction of IR counter-terms, we explore an alternative idea based in the application of the loop-tree duality (LTD) \cite{Catani:2008xa,Rodrigo:2008fp,Bierenbaum:2010cy,Bierenbaum:2012th,Bierenbaum:2013nja,Buchta:2014dfa,Buchta:2014fva,Buchta:2015xda,Buchta:2015jea,Buchta:2015lva,Buchta:2015wna,Hernandez-Pinto:2015ysa,Sborlini:2015uia,Sborlini:2016fcj}. The LTD theorem establishes that loop scattering amplitudes can be expressed as a sum of PS integrals (i.e. the so-called 
\textit{dual integrals}) with an additional physical particle. 
Dual integrals and real-radiation contributions exhibit a similar structure, and can be combined at integrand level. As shown in 
Refs.~\cite{Buchta:2014dfa,Hernandez-Pinto:2015ysa,Sborlini:2015uia,Sborlini:2016fcj}, the divergent behaviour of both contributions 
is matched, and the combined expression is finite. 
In other words, working in the context of dimensional regularisation (DREG)~\cite{Bollini:1972ui, 'tHooft:1972fi, Cicuta:1972jf, Ashmore:1972uj}
with $d=4-2\ep$ the number of space-time dimensions, the mapped real-virtual contributions do not lead to $\ep$-poles, 
which implies that the limit $\ep \to 0$ can safely be considered. This fact has a strong implication: 
the possibility of carrying out purely four-dimensional implementations for any observable at NLO and higher-orders. 
Thus, the aim of this paper is to implement a novel algorithm for a four-dimensional regularisation of multi-leg physical 
cross-sections at NLO free of soft and final-state collinear subtractions. 

It is worth to mention that the idea of obtaining purely four-dimensional expressions to compute higher-order observables 
has been previously studied. For instance, it was proposed to apply momentum smearing \cite{Soper:1998ye, Soper:1999xk, Soper:2001hu, Kramer:2002cd} to combine real and virtual contributions, 
thus achieving a local cancellation of singularities. Other alternative methods consist in rewriting the standard IR/UV subtraction counter-terms in a local form, as discussed in Refs. \cite{Becker:2010ng, Becker:2012aqa}, or modifying the structure of the propagators (and the associated Feynman rules) \cite{Pittau:2012zd,Donati:2013iya,Fazio:2014xea} to regularise the singularities. Besides that, the numerical computation of virtual corrections has received a lot of attention 
in recent years~\cite{Passarino:2001wv, Ferroglia:2002mz, Nagy:2003qn, Nagy:2006xy, Anastasiou:2007qb, Moretti:2008jj, Gong:2008ww, Kilian:2009wy, Becker:2012nk, Becker:2012bi,Freitas:2016sty}. For these reasons, through the application of LTD, we will tackle both problems simultaneously; we will express virtual amplitudes as phase-space integrals and combine them with the real contributions, working directly at \emph{integrand} level. Moreover, physically interpretable results will emerge in a natural way.

The outline of this article is the following. In Section~\ref{sec:notationINTRO} we introduce and review the basis of LTD. 
As starting point, we describe in detail the unsubtracted implementation of NLO corrections with an scalar toy example. 
First, the IR singular structure of the scalar three-point function and the construction of the corresponding dual integrals 
is commented in Section~\ref{sec:triangleCUT}. Second, the mapping of momenta between real and virtual corrections 
is defined in Section~\ref{sec:triangleREAL} for the toy example, and our four-dimensional regularisation of 
soft and collinear singularities is presented with a strong physical motivation. 
Then, in Section~\ref{sec:renorm} we study the renormalisation of UV divergences at integrand level in the LTD framework. 
The discussion is focused on the treatment of scalar two-point functions, and we properly rewrite unintegrated dual UV 
counter-terms in a fully local way. After that, we carefully analyse the 
implementation of these techniques to the process $\gamma^* \to q \bar q (g)$ in Section~\ref{sec:gammatoqqbar}. 
We put special emphasis in the algorithmic construction of the integrands, and in the numerical implementation of the purely 
four-dimensional representation. In Section~\ref{sec:global}, we generalise the unsubtraction algorithm to multi-leg processes, and 
briefly comment about the extension of the algorithm to NNLO. 
Finally, we present the conclusions and discuss the future research directions in Section~\ref{sec:conclusions}.

\section{Review of the loop-tree duality}
\label{sec:notationINTRO}
\begin{figure}[htb]
\begin{center}
\includegraphics[width=0.3\textwidth]{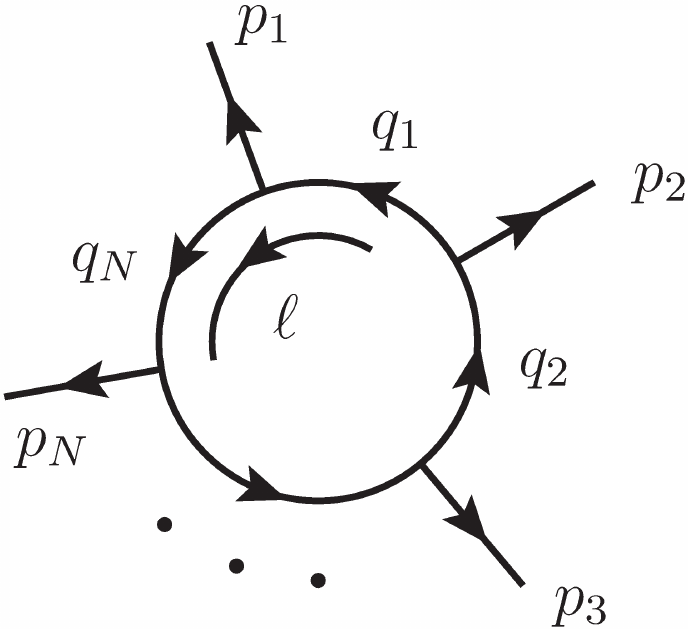}
\caption{Generic one-loop topology with $N$ external legs. 
All momenta are considered outgoing, and the internal momentum flow 
is taken counter-clockwise.
\label{fig:OneLoopScalar}}
\end{center}
\end{figure}

In this section we review the main ideas behind the LTD method.
The LTD theorem~\cite{Catani:2008xa} 
establishes a direct connection among loop and phase-space integrals. 
Explicitly, it demonstrates that loop contributions to scattering 
amplitudes in any relativistic, local and unitary quantum field theory 
can be computed through dual integrals, which are build from single cuts 
of the virtual diagrams. Let's consider a generic $N$-particle scalar one-loop integral, i.e. 
\beq
L^{(1)}(p_1, \dots, p_N) = \int_{\ell} \, 
\prod_{i \in \alpha_1} \,G_F(q_i)~,
\label{oneloop}
\eeq 
over Feynman propagators $G_F(q_i) = (q_i^2-m_i^2+\imath 0)^{-1}$,
whose most general topology is shown in Fig.~\ref{fig:OneLoopScalar}. Then, there is a corresponding dual representation consisting of the sum of 
$N$ dual integrals:
\beq
L^{(1)}(p_1, \dots, p_N) 
= - \sum_{i\in \alpha_1} \, \int_{\ell} \, \td{q_i} \,
\prod_{j \in \alpha_1, \, j\neq i} \,G_D(q_i;q_j)~, 
\label{oneloopduality}
\eeq 
where 
\beq
G_D(q_i;q_j) = \frac{1}{q_j^2 -m_j^2 - \imath 0 \, \eta \cdot k_{ji}}
\eeq
are dual propagators, and $i,j \in \alpha_1 = \{1,2,\ldots N\}$ label 
the available internal lines. In~\Eq{oneloop} and~\Eq{oneloopduality},
the masses and momenta of the internal lines are denoted $m_i$ and 
$q_{i,\mu} = (q_{i,0},\mathbf{q}_i)$, respectively, 
where $q_{i,0}$ is the energy and $\qb_{i}$ are 
the spatial components. In terms of the loop momentum $\ell$
and the outgoing four-momenta of the external particles $p_i$, the internal momenta are 
defined as 
\beq
q_{i} = \ell + k_i \ \ \ , \ \ \ k_{i} = p_{1} + \ldots + p_{i} \, ,
\eeq
together to the constraint $k_{N} = 0$ imposed by momentum conservation. 

On the other hand, the $d$-dimensional loop measure is given by
\beq
\int_{\ell} \bullet 
= - \imath \mu^{4-d} \int \frac{d^d \ell}{(2\pi)^{d}} \bullet~,
\eeq
and
\beq
\td{q_i} \equiv 2 \pi \, \imath \, \theta(q_{i,0}) \, \delta(q_i^2-m_i^2) \, ,
\label{eq:tddefinition}
\eeq
is used in \Eq{oneloopduality} to set the internal lines on-shell. Moreover, the presence of the Heaviside function restricts the integration 
domain to the positive energy region (i.e. $q_{i,0}>0$). Since LTD is derived through the application of the Cauchy's residue theorem, 
the remaining $d-1$ dimensional integration is performed over the forward on-shell hyperboloids defined by the solution of $G_F(q_i)^{-1}=0$ 
with $q_{i,0}>0$. Notice that these on-shell hyperboloids degenerate to light-cones when internal particles are massless.

The dual representation shown in \Eq{oneloopduality} is built by adding all possible single-cuts of the original loop diagram. In this procedure, 
the propagator associated with the cut line is replaced by \Eq{eq:tddefinition} whilst the remaining uncut Feynman propagators are promoted 
to dual ones. The introduction of dual propagators modifies the $\imath 0$-prescription since it depends on the sign of $\eta \cdot k_{ji}$, 
with $\eta$ a {\em future-like} vector, $\eta^2 \ge 0$, with positive definite energy $\eta_0 > 0$, and $k_{ji} = q_j - q_i$, which is independent 
of the loop momentum $\ell$ at one-loop. According to the derivation shown in Ref. \cite{Catani:2008xa}, 
$\eta$ is arbitrary so we can chose $\eta_\mu = (1,{\bf 0})$ to simplify the implementation. 

The difference between LTD and the Feynman Tree Theorem (FTT)~\cite{Feynman:1963ax,Feynman:1972mt}, where the loop integral 
is obtained after summing over all possible $m$-cuts, is codified in the dual prescription: correlations coming from multiple cuts in FTT 
are recovered in LTD by considering only single-cuts with the modified $\imath 0$-prescription. 
In other words, having different prescriptions for each cut is a necessary condition for the consistency of the method. 
As discussed in Ref.~\cite{Buchta:2014dfa}, the integrand in~\Eq{oneloopduality} becomes singular at the intersection 
of forward on-shell hyperboloids (FF case),  and forward with backward ($q_{j,0}<0$) on-shell hyperboloids (FB intersections). 
On one hand, the FF singularities cancel each other among different dual contributions; the change of sign in the modified 
prescription is crucial to enable this behaviour. On the other hand, the singularities associated with FB intersections remain 
constrained to a compact region of the loop-three momentum space and are easily reinterpreted in terms of causality. 
From a physical point of view, FB singularities take place when the on-shell virtual particle interacts with another 
on-shell virtual particle after the emission of outgoing on-shell radiation. The direction of the internal momentum flow establishes a natural causal ordering, and this interpretation is consistent with the Cutkosky rule. In fact, the total energy of the emitted particles, which is equal 
to $q_{i,0}-|q_{j,0}|$, has to be positive. Together with the positive energy constraint imposed by the delta distribution 
in~\Eq{eq:tddefinition}, it restricts the possible situations compatible with a sequential decay of on-shell physical particles.

\section{Singularities of the scalar three-point function}
\label{sec:triangleCUT}
In this section, we show a detailed derivation of the results presented in Ref.~\cite{Hernandez-Pinto:2015ysa}
concerning the scalar three-point function with massless internal particles. This discussion is useful 
to analyse and understand the application of LTD to the realistic case presented in Section~\ref{sec:gammatoqqbar}, 
and the posterior generalisation to multi-leg processes in Section~\ref{sec:global}.
We consider final-state massless and on-shell momenta labeled as $p_1$, $p_2$,
and the incoming momenta $p_3=p_1+p_2\equiv p_{12}$, by momentum conservation, with virtuality $p_3^2=s_{12}>0$. 
The internal momenta are $q_1 = \ell + p_1$,  $q_2 = \ell + p_{12}$ and $q_3 = \ell$, 
where $\ell$ is the loop momentum. The scalar three-point function at one-loop is given by 
the well-known result~\cite{Ellis:2007qk, Sborlini:2013jba}
\beqn
L^{(1)}(p_1,p_2,-p_3) &=& \int_\ell \, \prod_{i=1}^3 G_F(q_i) = - \cg \,
\frac{\mu^{2\ep}}{\ep^2} \, (- s_{12}-\imath 0)^{-1-\ep}~,
\label{triangle}
\eeqn
where $c_\Gamma$ is the usual loop volume factor (see \Eq{eq:cgamma} in Appendix~\ref{app:measure}). 
The LTD representation of the scalar integral in~\Eq{triangle}
consists of three contributions 
\beq
L^{(1)}(p_1,p_2,-p_3) = \sum_{i=1}^3 I_{i}~,
\label{DescomponeTrianguloTL}
\eeq
with 
\bea
I_1 &=& - \int_\ell \, \frac{\td{q_1}}{(2q_1\cdot p_2 - \imath 0)\,
(-2q_1\cdot p_1 + \imath 0)}~, \nn
\\ I_2 &=& - \int_\ell \, \frac{\td{q_2}  }{(-2q_2\cdot p_2 + \imath 0)\,  
(-2q_2\cdot p_{12} + s_{12} + \imath 0)}~,  \nn
\\ I_3 &=& - \int_\ell \, \frac{\td{q_3}}{(2q_3\cdot p_1 - \imath 0)\, 
(2q_3\cdot p_{12} + s_{12} - \imath 0)}~.
\label{CorteI3}
\eea
In order to simplify the computation of these integrals, 
we work in the centre-of-mass frame of $p_1$ and $p_2$,
and parametrise the momenta as
\beqn
p_1^{\mu} &=& \frac{\sqrt{s_{12}}}{2} \left(1, {\bf 0}_\perp, 1 \right)~, \quad
p_2^{\mu} = \frac{\sqrt{s_{12}}}{2} \left(1, {\bf 0}_\perp, -1 \right)~, \nn
\\ q_i^{\mu} &=& \frac{\sqrt{s_{12}}}{2} \, \xi_{i,0} \, 
\left(1, 2 \sqrt{v_i(1-v_i)} \, {\bf e}_{i,\perp}, 1-2v_i \right)~,
\label{Trianguloqiparametrizacion}
\eeqn
with $\xi_{i,0} \in [0,\infty)$ and $v_i \in [0,1]$ 
the integration variables describing the energy and polar angle of the loop momenta, respectively. 
Integration of the loop momentum in the transverse plane, which is described by the unit vectors 
${\bf e}_{i,\perp}$, is trivial in this case. 
The scalar products of internal with external momenta are given by
\beqn
2 q_i\cdot p_1/s_{12} &=& \xi_{i,0}\, v_i~, \nn
\label{qip1}
\\ 2 q_i\cdot p_2/s_{12} &=& \xi_{i,0}\, (1-v_i)~,
\label{qip2}
\eeqn
and the dual integrals in~\Eq{CorteI3} are rewritten as
\bea
I_1 &=& \frac{1}{s_{12}} \, \int d[\xi_{1,0}] \, d[v_1] \,  
\xi_{1,0}^{-1} \, (v_1 (1-v_1))^{-1}~, \nn
\\ I_2 &=& \frac{1}{s_{12}} \, \int d[\xi_{2,0}] \, d[v_2] \,  
\frac{(1-v_2)^{-1}}{1 - \xi_{2,0} + \imath 0}~, \nn
\\ I_3 &=& - \frac{1}{s_{12}} \, \int d[\xi_{3,0}] \, d[v_3] \,  
\frac{v_3^{-1}}{1 + \xi_{3,0}}~,
\label{eq:duals}
\eea
with the integration measure in DREG given by the direct product of
\beq
d[\xi_{i,0}] =  \frac{(4\pi)^{\ep-2}}{\Gamma(1-\ep)} \, \left(\frac{s_{12}}{\mu^2}\right)^{-\ep}
 \,  \xi_{i,0}^{-2\ep} \, d\xi_{i,0} \, , \qquad
d[v_i] =  \left(v_i(1-v_i)\right)^{-\ep} \, dv_i~.
\label{eq:measureDREG}
\eeq
It is possible to perform these integrals analytically; the result is
\bea
I_1 &=& 0~, \nn
\\ I_2 &=& \cgt \, \frac{\mu^{2\ep}}{\ep^2} \, s_{12}^{-1-\ep} \, e^{i 2\pi \ep}~, \nn
\\ I_3 &=& \cgt \, \frac{\mu^{2\ep}}{\ep^2} \, s_{12}^{-1-\ep}~, 
\label{eq:dualesresultados}
\eea
where the phase-space volume factor $\cgt$ is defined in~\Eq{eq:cgamma}.
As expected, the sum of the three dual integrals agrees with the well-known result 
from~\Eq{triangle}. It is worth noticing here that the integrand of $I_1$ is both IR and UV divergent. However, the application of DREG leads to equal and opposite $\ep$-poles, which justifies the result shown in \Eq{eq:dualesresultados}. We will discuss this fact more carefully in Sec. \ref{ssec:masslessbubbles}, since it plays a crucial role in the whole implementation of the LTD approach for physical processes.

Notice that in~\Eq{eq:duals} the dual $+\imath 0$ prescription is crucial for computing $I_2$, 
because $1-\xi_{2,0}$ changes sign inside the integration region, leading to a threshold singularity. For later use, it will be necessary an explicit expression for the imaginary part of $I_2$ at the integrand level, which is determined by setting the integrand 
of $I_2$ on the negative-energy on-shell mode of $G_F(q_3)$, i.e.
\bea
&& \imath \, {\rm Im} \, L^{(1)}(p_1, p_2, -p_3) = \imath \, {\rm Im} \, I_2 = 
\frac{1}{2} \, \int_\ell \, G_D(q_2; q_1) \, \td{q_2} \, \td{-q_3} \nn \\ && =
- \frac{\imath\, \pi}{s_{12}} \, \int d[\xi_{2,0}] \, 
d[v_2] \,  (1-v_2)^{-1} \, \theta(2-\xi_{2,0}) \, \delta(1 - \xi_{2,0}) 
= \imath \, \cgt \, \frac{\mu^{2\ep}}{2 \ep^2} \, s_{12}^{-1-\ep}\, \sin(2\pi \ep)~.
\label{eq:imI2}
\eea
We remark that $G_D(q_2; q_1) = G_F(q_1)$ because $\eta\cdot k_{12} = - p_{2,0} < 0$ with $\eta_\mu=(1,\bf{0})$,
and hence~\Eq{eq:imI2} is consistent with the Cutkosky's rule, which is determined by setting two propagators 
on-shell and outgoing, namely, by reversing the momentum flow of one of the on-shell internal lines~\cite{Catani:2008xa}. 
This is the causality connection mentioned in Section~\ref{sec:notationINTRO}, and it becomes relevant in our 
computation because the $\epsilon$-expansion of~\Eq{eq:imI2} reveals the presence of a purely imaginary single-pole 
in $I_2$ that will not be cancelled by real corrections. 
At integrand level, this means that the real part of $I_2$ presents an integrable singularity in the neighbourhood 
of $\xid=1$, but there is also a non-integrable singularity that must be cancelled by removing its
imaginary component before performing a four-dimensional numerical implementation. 
Thus, the real part of $I_2$ is defined as 
\beqn
\re \, I_2 = I_2 - \imath \, \im \, I_2 &=& \frac{1}{s_{12}} \, \int d[\xi_{2,0}] \, d[v_2] \,  
(1-v_2)^{-1} \left(\frac{1}{1 - \xi_{2,0} + \imath 0}+ \imath \pi\delta(1-\xi_{2,0}) \right) \, ,
\eeqn 
and, by virtue of the Sokhotski–-Plemelj theorem,
\beqn
\re \, I_2 &=& \frac{1}{s_{12}} \, \int d[\xi_{2,0}] \, d[v_2] \,  
(1-v_2)^{-1} \, {\rm PV}\left(\frac{1}{1 - \xi_{2,0}}\right) \, ,
\eeqn 
where we make use of the Cauchy's principal value (PV) to get rid of the $+\imath 0$ prescription and the imaginary pole. From the formal point of view, we could perform this computation by simply working with the real part of the integrand (and neglecting the prescription). However, numerical instabilities arise and the application of PV prescription leads to a more efficient implementation.

\begin{figure}[htb]
\begin{center}
\includegraphics[width=0.3\textwidth]{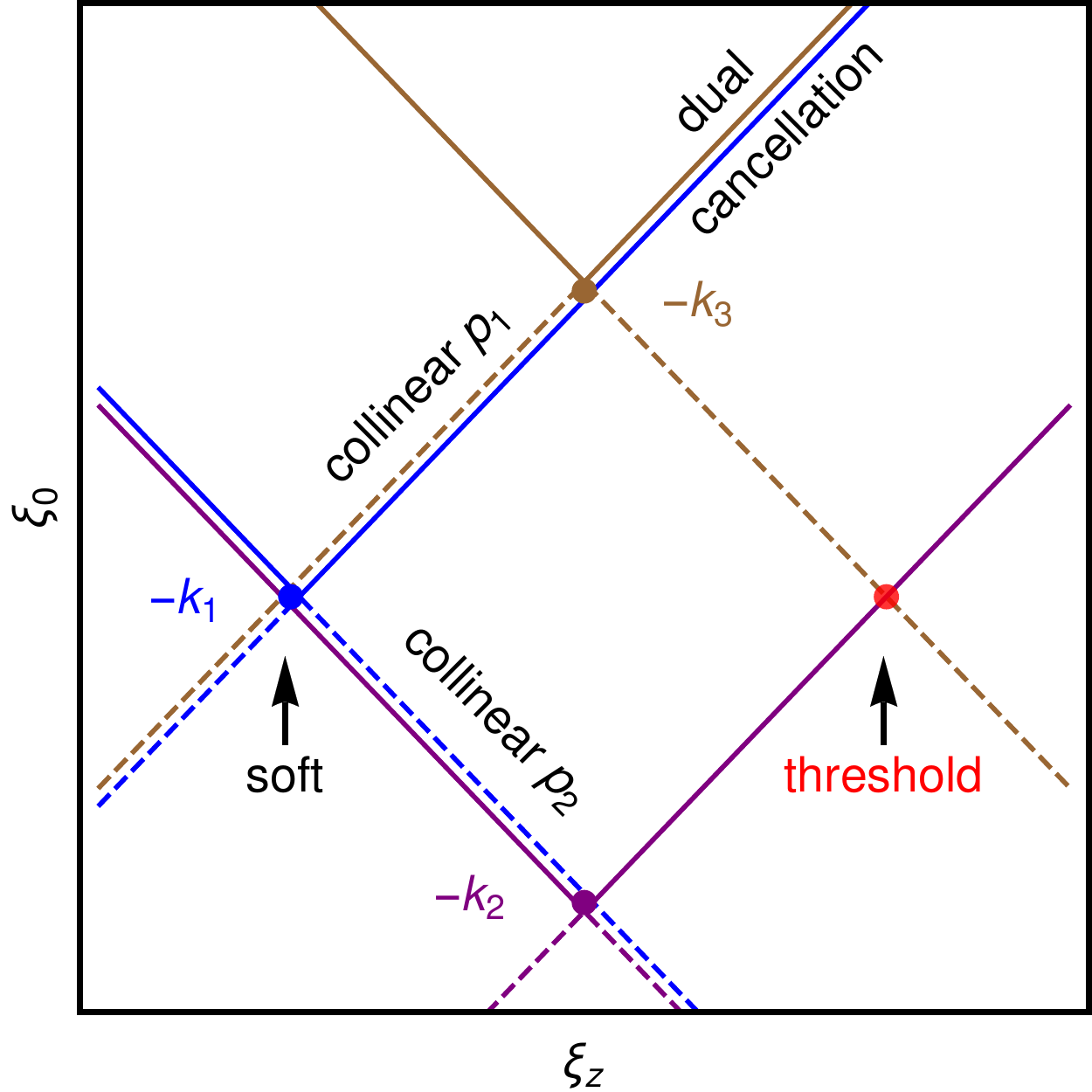} $\quad$
\includegraphics[width=0.3\textwidth]{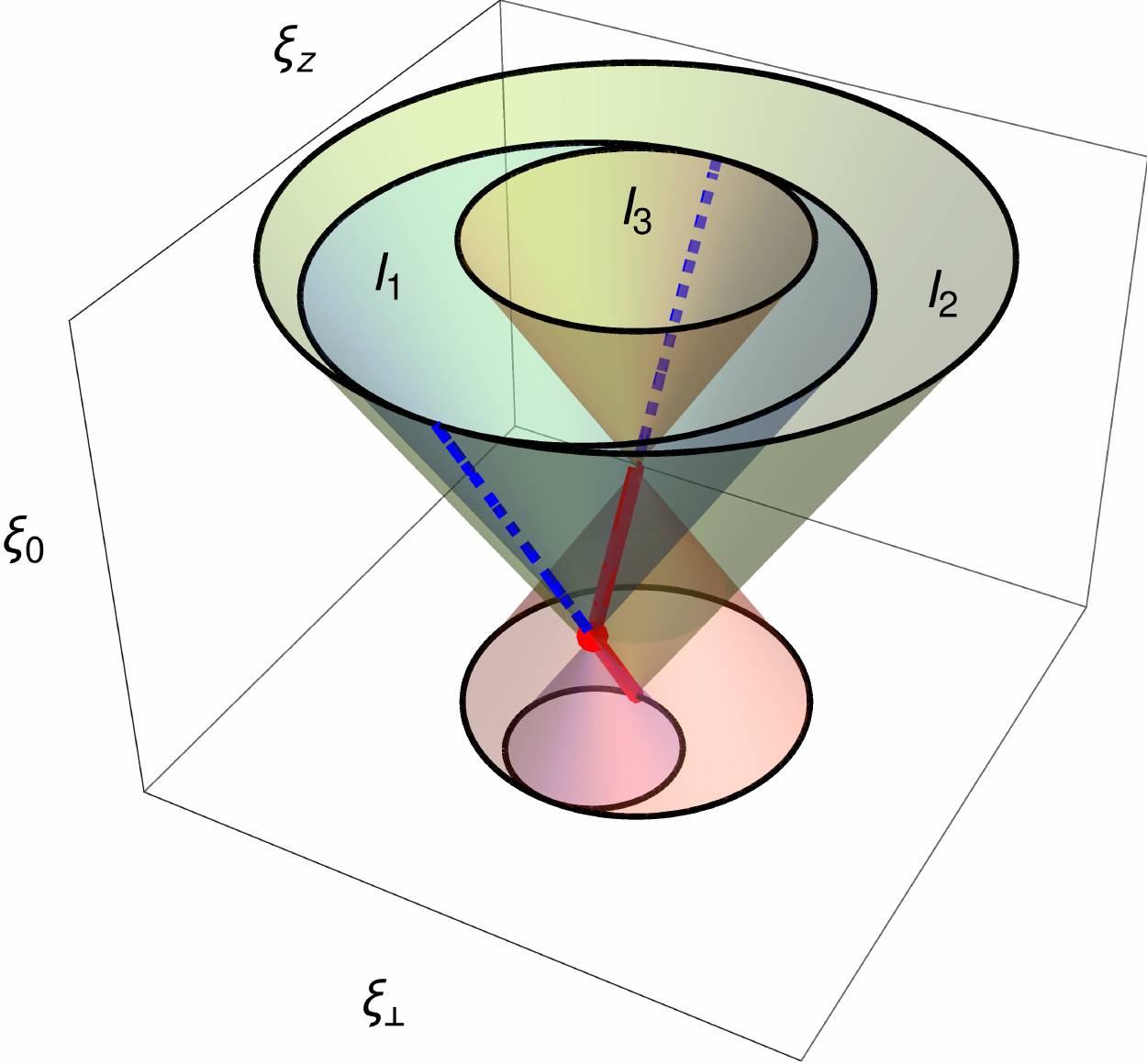} $\quad$
\includegraphics[width=0.3\textwidth]{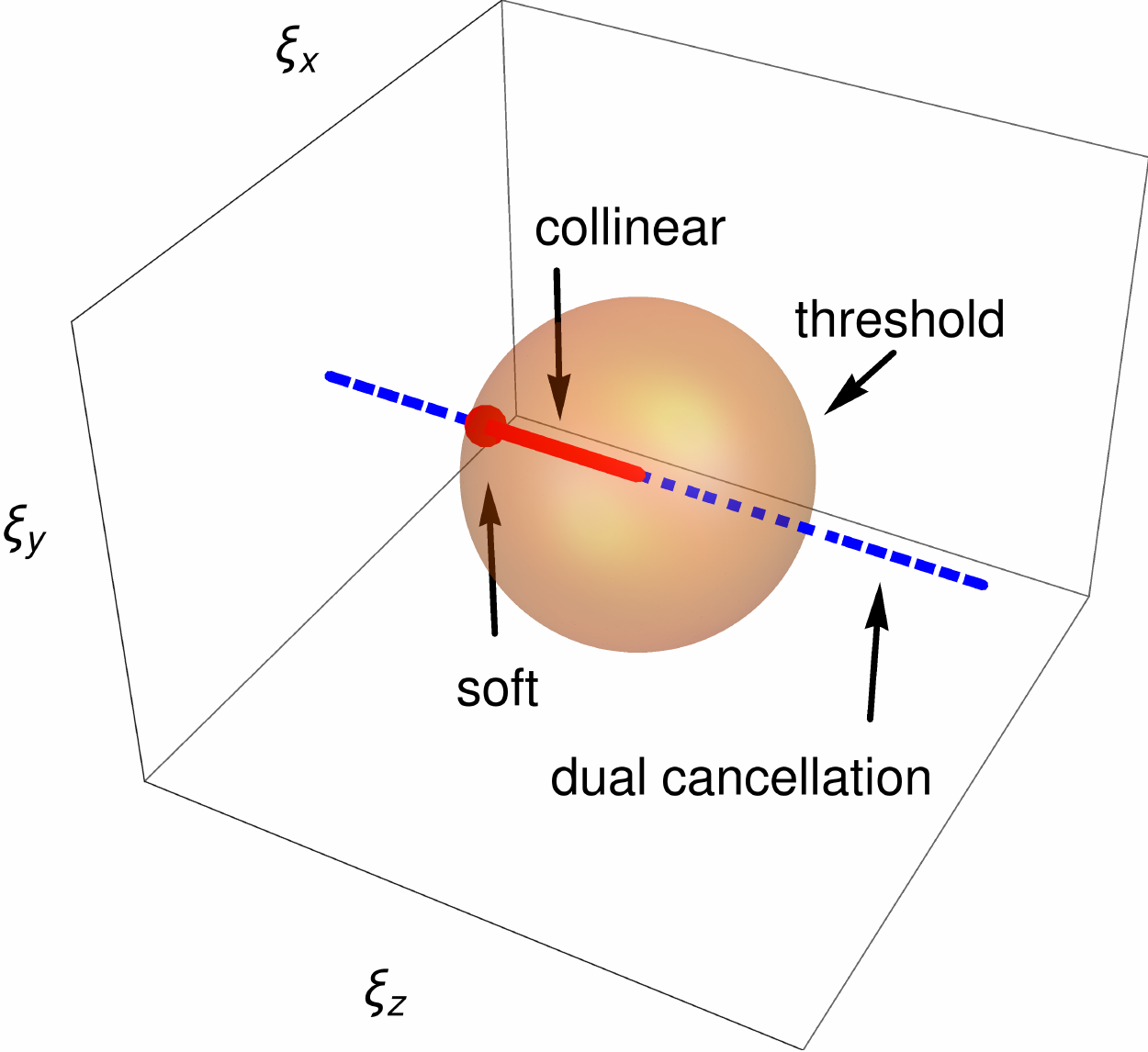}
\caption{Light-cones of the three-point function in the loop coordinates $\ell^{\mu}= \sqrt{s_{12}}/2\, (\xi_0,\xi_\perp,\xi_z)$,
with $\xi_\perp = \sqrt{\xi_x^2+\xi_y^2}$;
two dimensions (left) and three-dimensions (middle). LTD is equivalent
to integrate along the forward-light cones (solid lines in the left plot). 
Backward light-cones are represented by dashed lines.
The intersection of light-cones leads to soft, collinear, and threshold singularities
in the loop three-momentum space (right plot), or to the cancellation of singularities 
among dual contributions. 
\label{fig:REGIONLightcone}}
\end{center}
\end{figure}

LTD can further be exploited to have a deeper and detailed understanding of the origin 
of the singularities of the loop integral under consideration. As commented before, the origin of the 
singularities can be underlined by analysing the relative position and intersections of the on-shell 
hyperboloids or light-cones of the propagators entering the loop integrand~\cite{Buchta:2014dfa}. 
In Fig.~\ref{fig:REGIONLightcone}, we plot the light-cones that support each of the dual integrals of 
the dual representation of the three-point function in~\Eq{DescomponeTrianguloTL}. 
Although the scalar three-point function is UV finite, the individual dual integrals in~\Eq{CorteI3} diverge 
in the UV because dual propagators are linear in the loop momentum. 
However, their sum has the same UV singularities present in the original integral, as expected 
according to the LTD theorem. Thus, we can focus on its IR behaviour; 
renormalisation of UV divergences will be considered later in Section~\ref{sec:renorm}.
 
Collinear divergences are associated with an extended region supported in the intersection of light-cones, 
as shown in Fig.~\ref{fig:REGIONLightcone}. At large loop momentum, the intersections 
occur among forward light-cones and the collinear singularities cancel in the sum of dual integrals. 
However, there are still collinear divergences originated in the compact region defined by the intersection 
of forward and backward light-cones. Soft singularities arise at $q_{i,0}^{(+)}=0$ (point-like solution), 
but the soft singularities of the integrand at $q_{i,0}^{(+)}=0$ lead 
to divergences only if two other propagators -- each one contributing
with one power in the infrared -- are light-like separated from the $i^{th}$ propagator. 
In Fig.~\ref{fig:REGIONLightcone} this condition is fulfilled only at $q_{1,0}^{(+)}=0$. 
Finally, a threshold singularity appears at the dual integral $I_2$ through the intersection of the backward light-cone 
of $G_F(q_3)$ with the forward light-cone of $G_F(q_2)$. The imaginary part of $I_2$ is singular 
but this singularity can be removed by using \Eq{eq:imI2}. The singularity of the real part
of $I_2$ is integrable but will lead to instabilities by a direct numerical 
computation. In that case, a contour deformation is employed to achieve an stable 
numerical implementation~\cite{Buchta:2015xda,Buchta:2015jea,Buchta:2015lva,Buchta:2015wna}.

Motivated by Fig.~\ref{fig:REGIONLightcone}
and in order to isolate the IR divergences of the scalar three-point function, 
we define the soft and collinear components of the dual integrals 
in well-defined compact regions of the loop three-momentum, i.e. 
\bea
I_1^{\rm (s)} &=& I_1(\xi_{1,0}\le w)~, \nn
\label{I1softORIGINAL}
\\ I_1^{\rm (c)} &=&  I_1(w \le \xi_{1,0}\le 1~; v_1\le1/2)~, \nn
\label{I1collinealORIGINAL}
\\ I_2^{\rm (c)} &=& I_2(\xi_{2,0}\le 1+w~; v_2\ge1/2)~,
\label{I2collinealORIGINAL}
\eea
where $0<w<1$ is a cut in the energy of the internal on-shell particle. 
The collinear singularity of the dual integral $I_2$ appears at $v_2=1$ with $\xid \in [0,1]$, but 
$I_2$ develops also a threshold singularity at $\xid=1$. For that reason we have imposed a finite 
$w$-cut to include the threshold region in the definition of $I_2^{(c)}$.
The integral $I_1^{\rm (s)}$ includes the soft singularity of the dual integral $I_1$ at $\xiu=0$, 
and the collinear singularities in the neighbourhood of $\xiu=0$ at $v_1=0$. The 
$\ep$-poles present in the integral $I_1^{\rm (c)}$ are due to collinear singularities only.
There is some arbitrariness in the definition of the 
integration regions of these integrals. Provided that we include the soft and collinear 
singularities, different definitions will differ only in the finite contributions. Indeed, we will 
redefine them later with a better motivated physical target.
These are, however, the simplest choice for the current illustrative purpose. 
Analytic integration gives 
\beqn
I_1^{\rm (s)} &=& \cgt \, \frac{w^{-2\ep}}{\ep^2} \, \mu^{2\ep} \, s_{12}^{-1-\ep} 
\, \frac{\sin (2\pi \ep)}{2\pi \ep}~, \nn
\\ I_1^{\rm (c)} &=& \cgt \,  \frac{1 - w^{-2\ep}}{2\ep^2} \, \mu^{2\ep} \, s_{12}^{-1-\ep} 
\, \frac{\sin (2\pi \ep)}{2\pi \ep}~, \nn
\\ \nn I_2^{\rm (c)} &=& - \cgt  \, \frac{(1 + w)^{1-2\ep}}{2(1-2\ep)\ep}
\, \left(1+\frac{4^{\ep} \Gamma(1-2\ep)}{\Gamma^2(1-\ep)} \right) \,
{}_2F_1\left(1, 1-2\ep, 2-2\ep; \frac{s_{12}(1+w)}{s_{12}+\imath 0}\right)  
\\ &\times& \, \mu^{2\ep} \, s_{12}^{-1-\ep} \, \frac{\sin (2\pi \ep)}{2\pi \ep}~.
\label{eq:collinearandsoft}
\eeqn
Using Pfaff and shift identities, the hypergeometric function in $I_2^{\rm (c)}$ 
can be written in the physical region with $w>0$ and $s_{12}>0$. This leads to
\beqn
\nn I_2^{\rm (c)} &=&  \cgt \, \frac{\mu^{2\ep}}{4\ep^2} \, s_{12}^{-1-\ep} \,
\left(1+\frac{4^{\ep} \Gamma(1-2\ep)}{\Gamma^2(1-\ep)} \right) \,
\\ &\times& \left[ e^{\imath 2 \pi \ep} - w^{-2\ep} \, 
{}_2F_1\left(2\ep, 2\ep, 1+2\ep; - \frac{1}{w}\right) 
\, \frac{\sin (2\pi \ep)}{2\pi \ep} \right]~, 
\label{eq:I2collinearA}
\eeqn
with 
\beq
{}_2F_1\left(2\ep, 2\ep, 1+2\ep; z\right) = 1+ 4\ep^2 \, {\rm Li}_2(z)
+ {\cal O}(\ep^3)~.
\eeq
As expected, the soft integral in~\Eq{eq:collinearandsoft} contains double poles, 
while the collinear integrals develop single poles only. 
Although each integral depends on the cut $w$,  the poles of the sum are 
independent of $w$ and agree with the total divergences of the full integral, 
i.e. $L^{(1)}(p_1,p_2,-p_3) = I^{\rm IR} + {\cal O}(\ep^0)$ with 
\bea
\nn I^{\rm IR} &=& I_1^{\rm (s)} + I_1^{\rm (c)} + I_2^{\rm (c)} =  
\, \frac{\cg }{s_{12}} \, \left(\frac{-s_{12}-\imath 0}{\mu^2}\right)^{-\ep} 
\\ &\times&\left[\frac{1}{\ep^2}+\ln{2}\ln{w}-\frac{\pi^2}{3}
-2\li{-\frac{1}{w}}+\imath \pi \ln{2}\right] + {\cal O}(\ep)~.
\label{IIRexpansion}
\eea

Outside the region that contains the IR poles, the sum of the dual integrals is finite, although they are separately divergent. 
A suitable combination is required to obtain finite results. So, we define the forward and the backward regions as those contained 
in $v_i \le 1/2$ and $v_i\ge 1/2$, respectively. This separation 
does not have any physical meaning; it is just convenient for 
the analytic computation. Explicitly, we define 
\bea
I^{(\f)} &=& I_1(\xi_{1,0}\ge 1~; v_1\le1/2)  
+ I_2(v_2\le1/2) + I_3(v_3\le1/2) \nn \\
&=& \frac{\cg}{s_{12}} \, \int_{0}^\infty d\xi_{0} \, \int_0^{1/2} \, dv \, 
\bigg[ \frac{1}{1 + \xi_{0}} \bigg(  
(1-v)^{-1} + 2 \, \ln{\frac{1 + \xi_{0}}{\xi_0}} \, \delta(v) \bigg) \nn \\ 
&+& \frac{(1-v)^{-1}}{1 - \xi_{0}+ \imath 0} \bigg]  + {\cal O}(\ep)~, 
\label{eq:forward} 
\eea
where we have performed a trivial change of the integration variables, i.e.
\beq
\xi_{1,0} = 1+\xi_0~ , \quad \xi_{2,0} = \xi_{3,0} = \xi_0~, \quad v_i = v~,
\label{eq:cdvforward}
\eeq
and we have taken the limit $\epsilon \to 0$ at integrand level. Notice that each dual integrand is still individually singular. 
For instance, $I_1$ and $I_3$ are divergent at $v_1 = 0 = v_3$ but their sum is finite in this limit, although UV divergences survive in the sum. 
The divergent high-energy behaviour of $I_1+I_3$ is cancelled once we add the contribution of $I_2$.
These cross-cancellations of singularities allow to perform the integral of the forward contribution with $\ep=0$. 
The logarithmic terms in~\Eq{eq:forward} are originated from the fact that we are using different coordinate systems for each 
dual integral. This produces a mismatch of the integration measures that is of ${\cal O}(\epsilon)$. 
Since the integral behaves as ${\cal O}(\epsilon^{-1})$ in the collinear limit, a non-vanishing
finite contribution arises from the collinear region. 
Explicitly, the expansion of $v^{-1-\epsilon}$ by using \Eq{Distro1} leads to 
\beqn
\nn && \left(-\frac{1}{\epsilon} \, \delta(v) + \left(\frac{1}{v}\right)_{C}+{\cal O}(\epsilon) \right) 
\left( (1+\xi_0)^{-2\epsilon}\, (1-v)^{-1}-\xi_0^{-2\epsilon}\right) \\
&& =  (1-v)^{-1} + 2 \, \ln{\frac{1+\xi_0}{\xi_0}} \delta(v) + {\cal O}(\epsilon) \, ,
\label{eq:logsinforward}
\eeqn 
where we removed the $C$-distribution. These logarithms are avoidable by a proper reparametrisation of the integration 
variables; we will return to this point in Section~\ref{ssec:unify}. From~\Eq{eq:forward}, we obtain
\beq
I^{(\f)} = \frac{\cg}{s_{12}} \, 
\left[ \frac{\pi^2}{3} - \imath  \pi \, \ln{2} \right] + {\cal O}(\ep)~.
\label{IntegralFORWARD}
\eeq

In an analogous way, we can compute the finite contribution originated in the backward region ($v_i \ge 1/2$), which is given by 
\bea
\nn I^{(\b)} &=&  I_1(\xi_{1,0}\ge w~; v_1\ge1/2) 
+ I_2(\xi_{2,0}\ge 1+w~; v_2\ge1/2) + I_3 (v_3\ge 1/2)
\\ \nn &=& \frac{\cg}{s_{12}}
\int_{0}^\infty d\xi_{0} \, \int_{1/2}^1  \, dv \,  
\bigg[ \frac{1}{w + \xi_0} \left( v^{-1} + 2 \, \ln{\frac{w + \xi_0}{1 + w + \xi_0}}  
\, \delta(1-v) \right) 
\\ &-& \frac{v^{-1}}{1 + \xi_0} \bigg]  + \, {\cal O}(\ep)~,
\label{eq:backward}
\eea
where we throw the $+\imath 0$ prescription because $w>0$ excludes the threshold singularity 
from the integration region. Also, to obtain \Eq{eq:backward} starting from \Eq{eq:duals}, we used the 
following change of variables
\beq
\xi_{1,0} = w+\xi_0~ , \quad \xi_{2,0} = 1+\xi_0 +w~, \quad \xi_{3,0} = \xi_0~, \quad v_i = v,
\label{eq:cdvbackward}
\eeq
and then we took the limit $\epsilon \to 0$ at integrand level. Similarly to the forward integral, 
there is a cancellation of collinear singularities among $I_1$ and $I_2$, which takes place at $v_1=1=v_2$ in this case; 
thus $I_1+I_2$ is IR-finite but it is still UV-divergent. To regularise the high-energy behaviour, we need to add 
the contribution of $I_3$. Again, notice that we have introduced some logarithmic terms in \Eq{eq:backward}. 
These terms are due to the mismatch in the collinear behaviour of $I_1$ and $I_2$ at ${\cal O}(\ep)$. 
As we did for the forward case, we derived the logarithmic corrections by expanding the collinear factor 
$(1-v)^{-1-\ep}$ and the integration measure after the implementation of~\Eq{eq:cdvbackward}; i.e.
\beqn
\nn && \left(-\frac{1}{\epsilon} \, \delta(1-v) + \left(\frac{1}{1-v}\right)_{C}+{\cal O}(\epsilon) \right)
 \left( (w+\xi_0)^{-2\epsilon}\, v^{-1} - (1+w+\xi_0)^{-2\epsilon}\right) 
\\ && = v^{-1} + 2 \, \ln{\frac{w+\xi_0}{1+w+\xi_0}} \delta(1-v) + {\cal O}(\epsilon) \, ,
\label{eq:logsinbackward}
\eeqn
and, again, the $C$-distribution can be removed because the integrand is regular for $v=1$. 
So, the integral over the sum of the three dual integrands can be performed with $\ep = 0$, and we obtain
\beq
I^{(\b)} = \frac{\cg}{s_{12}} \, \left[2
{\rm Li}_2\left(-\frac{1}{w}\right) - \ln{2} \ln{w} \right] + {\cal O}(\ep)~.
\label{IntegralBACKWARD}
\eeq
The sum of \Eq{IIRexpansion}, \Eq{IntegralFORWARD} and \Eq{IntegralBACKWARD}
leads to the correct full result up to ${\cal O}(\ep)$
\beq
L^{(1)}(p_1,p_2,-p_3) = I^{\rm IR} + I^{(\b)} + I^{(\f)} + {\cal O}(\ep)~,
\eeq
which is independent of $w$. This was expected because $w$ is a non-physical cut. 
It is worth noting that only $I^{\rm IR}$ contains the $\ep$-poles and the remaining contributions have been computed directly with $\ep=0$. Moreover, through the application of Eqs. (\ref{eq:logsinforward}) and (\ref{eq:logsinbackward}), the integrand can easily be expressed as the 
$\epsilon \to 0$ limit of the original DREG expression plus some logarithmic corrections which lead to the right result. 

\subsection{Unifying the coordinate system}
\label{ssec:unify}
As a final remark, we point out that it is possible to avoid the introduction of the extra logarithmic terms 
in $I^{\rm (f)}$ and $I^{\rm (b)}$. They are originated from the fact that each dual integrand 
has been expressed in a different coordinate system and they approach the collinear limit in a slightly different way 
at ${\cal O}(\epsilon)$.  The solution consists in using the same coordinate system for all the dual integrals, 
i.e. we have to map exactly the loop three-momenta $\qb_i$. Although for analytic computations this leads to more complex 
integration limits, for numerical applications it is indeed the natural choice for the implementation of this approach. 
For instance, for the forward integral $I^{\rm (f)}$ the integration variables ($\xi_{1,0}$, $v_1$) must be written
in terms of ($\xi_{3,0}$, $v_3$). Notice that $q_3=\ell$ is set on-shell in $I_3$, but not in $I_1$ where $q_1^2=0$. 
From~\Eq{Trianguloqiparametrizacion}, the spatial components of the internal loop momenta are parametrised according to
\bea
\qb_1 &=& \frac{\sqrt{s_{12}}}{2} \, \xi_{1,0} \, 
\left(2 \sqrt{v_1(1-v_1)} \, {\bf e}_{1,\perp}, 1-2v_1 \right) \nn \\ 
&=& \qb_3 + \pb_1 = \frac{\sqrt{s_{12}}}{2} \,  
\left(\xi_{3,0} \, 2 \sqrt{v_3(1-v_3)} \, {\bf e}_{3,\perp}, \xi_{3,0} (1-2v_3)+1 \right) \, ,
\eeqn 
with $q_{1,0}^{(+)}=\sqrt{(\qb_3+\pb_1)^2-\imath 0}$ when $q_1$ is on-shell, which leads to 
\bea
\xi_{1,0} &=& \sqrt{(1+\xi_{3,0})^2 -4 v_3\, \xi_{3,0}}~,  \nn \\
v_1 &=& \frac{1}{2} \left(1 - \frac{1 + (1-2v_3)\, \xi_{3,0}}
{\sqrt{(1+\xi_{3,0})^2 -4 v_3\, \xi_{3,0}}} \right)~.
\label{eq:changexi10}
\eea
With this change of variables, we find the following representation of the forward integral with $\ep=0$,
\bea
\nn I^{(\f)} &=& \frac{\cg}{s_{12}} \, \bigg\{ \int_0^{1/2} \, dv \, \int_0^{\infty} d\xi_{0} \, \bigg[
v^{-1} \bigg( \frac{(1-v)^{-1}}{\sqrt{(1+\xi_0)^2-4v\, \xi_0}}
- \frac{1}{1+\xi_0} \bigg) + \frac{(1-v)^{-1}}{1-\xi_0+\imath 0} \bigg]
\\ && + \int_0^{1/2} \, dv_1 \, \int_1^{1/(1-2v_1)} d\xi_{1,0} \, 
\xi_{1,0}^{-1} \, (v_1(1-v_1))^{-1} \bigg\}  
+ {\cal O}(\ep)~,
\label{eq:forward2}
\eea
which is free of the logarithmic contributions that appear in \Eq{eq:forward}, and leads to the same result as in \Eq{IntegralFORWARD}. Notice that in \Eq{eq:forward2} we used $v_2=v_3=v$ and $\xid=\xit=\xi_0$. A similar representation is available for the backward integral, where we must combine $I_1$ and $I_2$, by expressing ($\xiu$, $v_1$) in terms of ($\xid$, $v_2$). Explicitly, we use
the same change of variables as in~\Eq{eq:changexi10}, which leads to
\bea
\nn I^{(\b)} &=& \frac{\cg}{s_{12}} \,  \int_{1/2}^1 \, dv \, \int_{0}^{\infty} d\xi_{0} \bigg[
(1-v)^{-1} \bigg( \frac{v^{-1} \, \theta\left(\sqrt{(1+\xi_{0})^2 -4 v\, \xi_{0}}-w\right)}{\sqrt{(1+\xi_0)^2-4v\, \xi_0}} 
\, \theta\left( \xi_0 - \frac{1}{2v-1}\right)
\\ &&  + \, \frac{\theta\left(\xi_{0}-1-w\right)}{1-\xi_0+\imath 0} \bigg) - \frac{v^{-1}}{1+\xi_0} \bigg] + {\cal O}(\ep)~,
\label{eq:backward2}
\eea
where we also applied $v_2=v_3=v$ and $\xi_{2,0}=\xi_{3,0}=\xi_0$.
The integration limits in~\Eq{eq:backward2}, codified through the Heaviside theta functions, are more cumbersome 
than in~\Eq{eq:backward}, but the result of both expressions is the same, and is given by~\Eq{IntegralBACKWARD}.
 
In summary, the IR singularities of loop integrals are restricted to a compact area 
of the integration domain, and the finite remnants are expressible in terms of pure four-dimensional functions, 
which implies that DREG could be avoided. Still, we need to keep $d\neq 4$ to deal with $I^{\rm IR}$; 
in the following we will show how to overcome this issue to achieve a fully four-dimensional implementation. 
 
\section{Unsubtraction of soft and collinear divergences}
\label{sec:triangleREAL}

In Section~\ref{sec:triangleCUT}, we have illustrated in detail the application of LTD to a scalar one-loop Feynman 
integral and we have isolated its infrared divergences in the function $I^{\rm IR}$ (see \Eq{IIRexpansion}), which is obtained 
from a compact region of the loop three-momentum. In the framework of LTD, a suitable mapping of external and loop momenta 
between virtual and real corrections allows to cancel the IR singularities at the integrand level, such that a full four-dimensional 
implementation is achieved without the necessity to introduce soft and collinear subtraction terms~\cite{Hernandez-Pinto:2015ysa}. 
We illustrate the method with a simplified toy scalar example before affording a complete calculation in a realistic physical process 
in Section~\ref{sec:gammatoqqbar}. 

We consider the one-loop virtual corrections to the cross-section that are proportional to the scalar three-point function 
\beqn
\sigma_{\rm V}^{(1)} &=& \frac{1}{2 s_{12}} 
\int d\Phi_{1\to 2} \, 2 \, \re \, \la {\cal M}^{(0)} | {\cal M}^{(1)} \ra 
= - \sigma^{(0)} \, 2 g^2 \, s_{12} \,  {\rm Re} \, L^{(1)}(p_1,p_2,-p_3)~,
\eeqn
where 
\beqn
\sigma^{(0)} &=& \frac{g^2}{2 s_{12}} \, \int d\Phi_{1\to 2}~,
\eeqn
is the Born cross-section, $\int d\Phi_{1 \to 2}$ is the integrated phase-space 
volume, given in~\Eq{Medida12INTEGRADA}, and $g$ is a generic coupling. 

On the other hand, it is necessary to include also the real radiation due to $1\to 3$ processes. 
The momenta configuration is $p_3 \to p_1' + p_2' + p_r'$, where we keep the same incoming momentum  
as in the $1 \to 2$ contributions, where $p_3\to p_1+p_2$ with $p_3^2 = s_{12}$, 
since we aim to obtain a local cancellation of singularities.
The real radiation correction to the cross-section is given by 
\beq
\sigma_{\r}^{(1)} = \frac{1}{2 s_{12}} \,  
\int d\Phi_{1\to 3} \, 2 \re \, \la {\cal M}^{(0)}_{2r}| {\cal M}^{(0)}_{1r} \ra
= \frac{g^4}{2\, s_{12}} \,  \int d\Phi_{1\to 3} \, \frac{2 s_{12}}{s_{1r}' \, s_{2r}'}~,
\label{eq:realscalar}
\eeq
with $s_{ir}'= (p_i'+p_r')^2$. The real corrections included in~\Eq{eq:realscalar} can be understood as the 
interference of the two scattering amplitudes corresponding to the emission of the real radiation from each of the 
outgoing particles. We do not take into account for the moment the squares of these amplitudes, which are proportional to $1/s_{ir}'^2$. 
They are topologically related to self-energy diagrams and will be considered explicitly in 
Section~\ref{sec:gammatoqqbar} for the physical process $\gamma^* \to q \bar q (g)$. 
For the current illustrative purpose it is enough to consider this interference. 

Then, we split the three-body phase-space to isolate the different IR singular regions. This strategy is a common practice 
in the context of subtraction methods \cite{Kunszt:1992tn,Frixione:1995ms}, because it allows to 
optimise the local cancellation of the collinear singularities at integrand level. Since there are three particles in the 
final state and the incoming one is off-shell, it is enough to separate the three-body phase-space 
into two pieces by making use of the identity 
\beq
1=\theta(y_{2r}'-y_{1r}') + \theta(y_{1r}'-y_{2r}') \, ,
\label{eq:PS3decomposition}
\eeq
which leads to the definitions
\beq
\widetilde{\sigma}_{\r,i}^{(1)} = \frac{1}{2 s_{12}} \,  
\int d\Phi_{1\to 3} \, 2 \re \, \la {\cal M}^{(0)}_{2r}| {\cal M}^{(0)}_{1r} \ra
\, \theta(y_{jr}'-y_{ir}')~, \qquad i,j \in \{ 1,2\}~,
\label{eq:separacionREALtoymodel}
\eeq
where $y_{ir}'= s_{ir}'/s_{12}$ are dimensionless variables. Analogously, we define the corresponding 
dual contributions to the virtual cross-section as 
\beq
\widetilde{\sigma}_{\v,i}^{(1)} = \frac{1}{2 s_{12}} \,  
\int d\Phi_{1\to 2} \, 2 \re \, \la {\cal M}^{(0)}| {\cal M}^{(1)}_i \ra
\, \theta(y_{jr}'-y_{ir}')~,
\eeq
with 
\beq
\la \M{0}|{\cal M}^{(1)}_i\ra = - g^4 \, s_{12} \, I_i \, ,
\label{eq:m1dual}
\eeq
the \emph{dual components} of the one-loop scattering 
amplitude according to the decomposition suggested 
in~\Eq{DescomponeTrianguloTL}. So, we claim that
\beqn
\widetilde{\sigma}_i^{(1)} &=& \widetilde{\sigma}_{\v,i}^{(1)} + \widetilde{\sigma}_{\r,i}^{(1)}~, 
\label{SigmaiCOMBINADO}
\eeqn
with $i\in \{1,2\}$, is finite in the limit $\ep \to 0$ and can be expressed using a purely four-dimensional representation. 
It is worth appreciating that the dual integral $I_3$ is not necessary to cancel the IR singularities present in the real corrections. 
This behaviour was expected from the analysis shown in Section \ref{sec:triangleCUT}, explicitly from Eqs. (\ref{eq:collinearandsoft}) 
and (\ref{IIRexpansion}). The dual integral $I_3$ does not lead to collinear divergences that are not cancelled by the other 
dual contributions, therefore it is not necessary to define the corresponding $\widetilde{\sigma}_3^{(1)}$. 
In fact, $I_3$ will solely contribute to the definition of the IR finite virtual remnant, formerly described in terms 
of the backward and forward integrals.

We now implement a mapping between the final-state momenta of the loop amplitudes $\{p_1,p_2\}$, 
the loop three-momentum $\boldsymbol{\ell}$, and the final-state momenta of the real amplitudes $\{p_1', p_2' , p_r' \}$.
Momentum conservation and on-shell constraints must be fulfilled by $p_i$ and $p_i'$, 
simultaneously. Hence, assuming $q_1$ on-shell and $q_3=q_1-p_1$ off-shell, we propose
\bea
&& p_r'^{\mu} = q_1^\mu~, \qquad p_1'^{\mu}  
= p_{1}^\mu - q_{1}^\mu + \alpha_1 \, p_2^\mu~, \nn \\ 
&& p_2'^{\mu} = (1-\alpha_1) \, p_2^{\mu}~,  \qquad 
\alpha_1 = \frac{q_3^2}{2q_3\cdot p_2}~, 
\label{eq:mapping1}
\eea
to perform the evaluation of the dual cross-section in~\Eq{SigmaiCOMBINADO}. This mapping has many interesting properties, 
which deserve to be discussed. In first place, momentum conservation is automatically fulfilled as $p_1'+p_2'+p_r'=p_1+p_2$~, 
and all the final-state momenta in~\Eq{eq:mapping1} are on-shell.  
Also, it is suitable to describe collinear configurations where $p_1 \parallel q_1$, 
which are reached for $\alpha_1 \to 0$; this is crucial to properly combine the divergent regions of the loop and real 
contributions to the cross-section and achieve a fully local regularisation. Notice that although we are restricted to 
$1\to 2$ and $1\to 3$ kinematics, the mapping in~\Eq{eq:mapping1} can easily be extended to processes with an 
arbitrary number of external particles (see Section~\ref{sec:global}). 

\begin{figure}[t]
\begin{center}
\includegraphics[width=7cm]{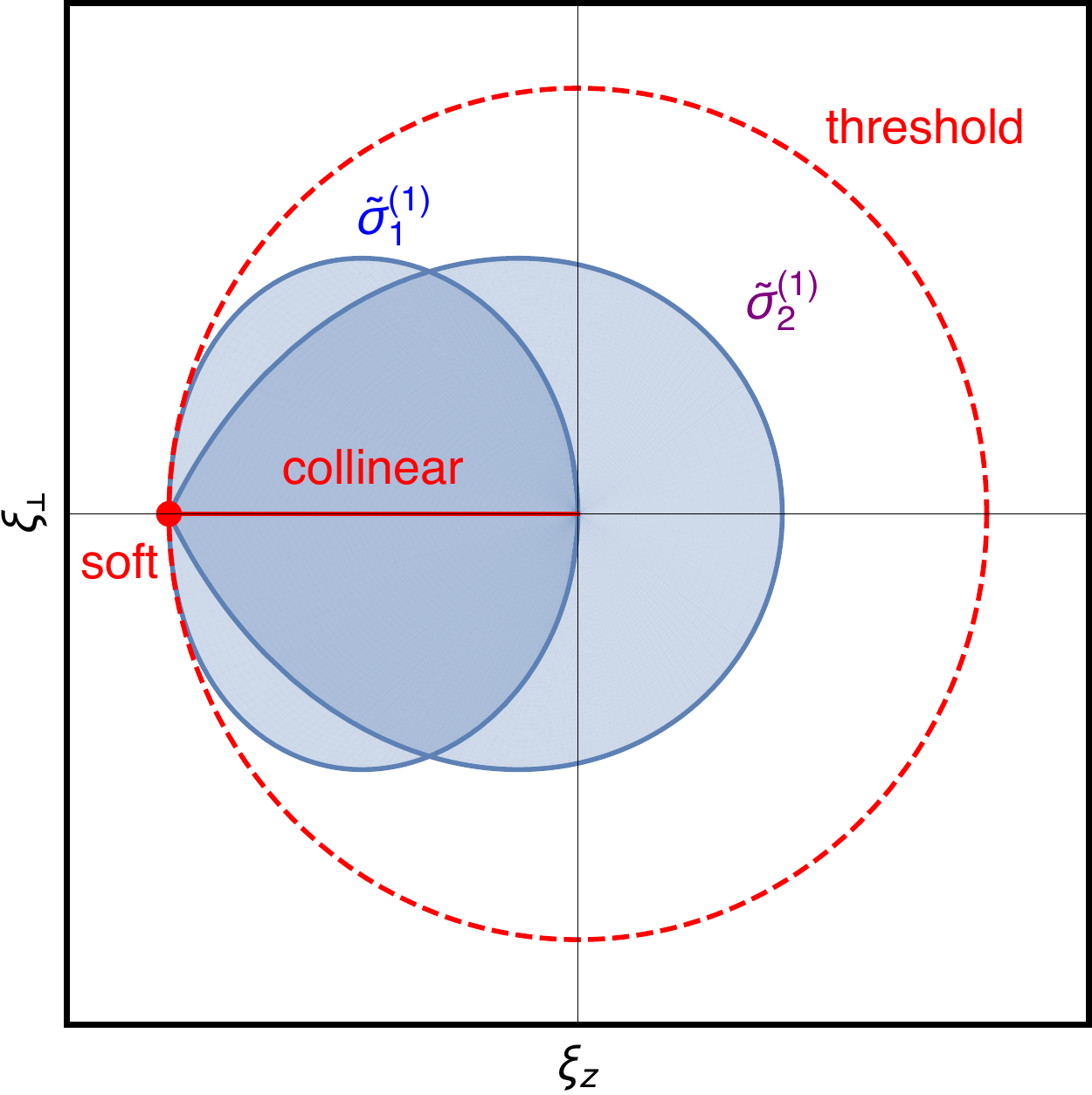}
\caption{\label{fig:dualregions} The dual integration regions in the loop three-momentum space.}
\end{center}
\end{figure}

The next step consists in using the parametrisation of $q_i$ and $p_i$ proposed 
in~\Eq{Trianguloqiparametrizacion}, together with the mapping in~\Eq{eq:mapping1}, 
to rewrite the two-body kinematic invariants $y_{ij}'$ in terms of the integration variables $(\xi_{1,0},v_1)$. 
Expressing the scalar products $p_i' \cdot p_j'$ with both sets of variables, we find
\beq
y_{1r}' = \frac{v_1 \, \xi_{1,0}}{1-(1 - v_1)\, \xi_{1,0}}~, \qquad 
y_{2r}' = \frac{(1-v_1)(1-\xi_{1,0}) \, \xi_{1,0}}{1 - (1-v_1)\,  \xi_{1,0}}~, \qquad
y_{12}' = 1-\xi_{1,0}~.
\label{eq:EcuacionYprima1}
\eeq
Since this mapping is optimised for the description of the collinear limit $p_1 \parallel q_1$, it must be used in the regions 
of the two-body and three-body phase-space where $y_{1r}'< y_{2r}'$~\footnote{To be mathematically rigourous, the transformation proposed in \Eq{eq:EcuacionYprima1} is a diffeomorphism connecting the physical three-body phase-space and its image in the 
integration domain of the dual contributions. Thus, in principle, it would not be necessary to define a new mapping, 
since it covers the whole phase-space in this simple example, although it is not really optimised for the collinear limit $p_2 \parallel q_1$.}.
The lower limit in the value of $y_{2r}'$ avoids to deal with $p_2 \parallel q_1$. Thus, a second mapping is necessary to treat the collinear limit with $y_{2r}'\to 0$, that is isolated in the region corresponding to $y_{2r}'< y_{1r}'$. With $q_2$ on-shell and $q_1=q_2-p_2$ off-shell, we define
\bea
&& p_2'^{\mu} = q_2^\mu~, \qquad 
p_r'^{\mu}  
= p_2^\mu - q_2^\mu + \alpha_2 \, p_1^\mu~,  \nn \\
&& p_1'^{\mu} = (1-\alpha_2) \, p_1^{\mu}~,   \qquad 
\alpha_2 = \frac{q_1^2}{2q_1\cdot p_1}~, 
\label{eq:mapping2}
\eea
and the two-body invariants are given by
\beq
y_{1r}' = 1-\xi_{2,0}~, \qquad 
y_{2r}' = \frac{(1-v_2) \, \xi_{2,0}}{1-v_2\, \xi_{2,0}}~, \qquad
y_{12}' = \frac{v_2 \, (1-\xi_{2,0}) \, \xi_{2,0}}{1-v_2 \, \xi_{2,0}}~.
\label{eq:EcuacionYprima2}
\eeq
By virtue of \Eq{eq:PS3decomposition}, the complete three-body phase space for the real radiation can be parametrised by applying \Eq{eq:mapping1} and \Eq{eq:mapping2}; in fact, it is useful to define 
\bea
\theta(y_{2r}'-y_{1r}') \equiv {\cal R}_1(\xiu, v_1) &=& \theta(1-2v_1) \, \theta\left(\frac{1-2v_1}{1-v_1}-\xiu\right)\, ,
\label{eq:mappingconditions1} 
\\ \theta(y_{1r}'-y_{2r}')\equiv {\cal R}_2(\xid, v_2) &=& \theta\left(\frac{1}{1+\sqrt{1-v_2}}-\xid\right) \, ,
\label{eq:mappingconditions2} 
\eea
to parametrise the integration regions for $\widetilde{\sigma}_{1}^{(1)}$ and $\widetilde{\sigma}_{2}^{(1)}$, respectively.
A graphical representation of the integration regions defined by Eqs.~(\ref{eq:mappingconditions1}) and (\ref{eq:mappingconditions2})
is shown in Fig.~\ref{fig:dualregions}.

Once the momenta are properly parametrised, we proceed to evaluate together the real and virtual contributions 
at integrand level. For $y_{1r}'< y_{2r}'$, we obtain
\bea
\label{eq:sigma1vr}
\widetilde  \sigma_1^{(1)} = \widetilde \sigma_{\v,1}^{(1)} + \widetilde \sigma_{\r,1}^{(1)} &=& 
\sigma^{(0)} \, 2 g^2 \, \int d[\xi_{1,0}] \, d[v_1] \, {\cal R}_1(\xiu, v_1) \, \nn \\ &\times&
\xi_{1,0}^{-1} \, (v_1 (1-v_1))^{-1} \,
\left[ \left(\frac{1-\xi_{1,0}}{1-(1-v_1)\, \xi_{1,0}} \right)^{-2\ep} - 1 \right]~, 
\eea
whilst
\bea
\widetilde \sigma_2^{(1)} = \widetilde \sigma_{\v,2}^{(1)} + \widetilde \sigma_{\r,2}^{(1)}  &=&
\sigma^{(0)} \, 2 g^2 \, \int d[\xi_{2,0}] \, d[v_2] \, {\cal R}_2(\xid,v_2) \nn \\ &\times& (1-v_2)^{-1} \,
\bigg[\frac{(1-\xi_{2,0})^{-2\ep}}{(1-v_2\, \xi_{2,0})^{1-2\ep}} 
- \frac{1}{1-\xi_{2,0}+\imath 0} - \imath \pi \delta(1-\xi_{2,0})\bigg]~, 
\label{eq:sigma2vr}
\eea
represents the analogous expression for $y_{2r}'< y_{1r}'$. The integrand in~\Eq{eq:sigma1vr} has the form 
\beq
\xi_{1,0}^{-1-2\ep} \, v_1^{-1-\ep} \, f(\xiu, v_1) \, , 
\eeq
and the function $f(\xiu, v_1)$ vanishes in the soft and collinear regions (i.e. $\xiu=0$ and/or $v_1=0$); 
moreover, $f(\xiu, v_1) = {\cal O}(\ep)$ which implies that
\beq
\widetilde \sigma_1^{(1)} = {\cal O} (\ep).
\label{eq:sigma1}
\eeq
On the other hand, the integrand of $\widetilde \sigma_2^{(1)}$ in~\Eq{eq:sigma2vr} 
behaves as $(1-v_2)^{-1-\ep} f(\xid, v_2)$, and $f(\xid, v_2)$ 
vanishes for $v_2=1$. The delta function in~\Eq{eq:sigma2vr} cancels the imaginary part of the $I_2$ dual integral, 
which is given by~\Eq{eq:imI2}. But indeed the condition $y_{2r}'<y_{1r}'$ excludes the threshold singularity 
of $I_2$ from the integration region with the exception of the single point at $(\xi_{2,0}, v_2) = (1, 1)$. 
This fact allows to calculate~\Eq{eq:sigma2vr} by removing the $+\imath 0$ prescription and the delta function. We obtain
\beq
\widetilde \sigma_2^{(1)} = 
- \sigma^{(0)} \, a \, \frac{\pi^2}{6}
+ {\cal O}(\ep)~,
\label{eq:sigma2}
\eeq 
with $a = g^2/(4\pi)^2$. It is important to notice that this result can be reached following two different paths. 
The first one consists in using DREG and integrating with $d=4-2\ep$; once an analytic expression is obtained, 
we verify that no $\epsilon$-poles are present and we take the limit $\ep \to 0$. On the other hand, we can consider directly 
the limit $\ep \to 0$ at integrand level; the expressions obtained are integrable and the results agree in both cases.

After combining virtual and real corrections, we define the virtual remnant $\overline \sigma^{(1)}_{\v}$ as the sum of the three dual 
integrals excluding the regions of the loop three-momentum already included in~\Eq{eq:sigma1vr} and~\Eq{eq:sigma2vr}:
\beqn
\nn \overline \sigma^{(1)}_{\v} &=& \sigma^{(0)} \, 2g^2 \, \int  d[\xi_0] \, d[v] \, 
\bigg[ - \left(1-{\cal R}_1(\xi_0,v)\right) \, \frac{v^{-1}(1-v)^{-1} }{\sqrt{(1+\xi_0)^2 - 4 v\, \xi_0}} 
\\ &-& \left(1-{\cal R}_2(\xi_0,v)\right) \, (1-v)^{-1}\,  \left(\frac{1}{1-\xi_0+\imath 0} + \imath \pi \delta(1-\xi_0)  \right) 
+ \frac{v^{-1}}{1+\xi_0}  \bigg]~.
\label{eq:remnant}
\eeqn
This expression is analogous to the sum of the forward and backward contributions defined in 
Section~\ref{sec:triangleCUT}, but does not require any unphysical cut $w$ to deal properly with the 
threshold singularity. In the previous expression, we have identified all the integration variables, 
$\xi_0=\xid=\xit$ and $v=v_2=v_3$, while $(\xiu,v_1)$ are expressed in terms of 
$(\xit,v_3)$  by using the change of variables in~\Eq{eq:changexi10} to directly avoid the appearance
of logarithmic contributions from the expansion of the integration measure. 
The integration regions are defined through
\bea
&&{\cal R}_1(\xi_0,v) = \left. \theta (1-2v_1) \, \theta\left( \frac{1-2v_1}{1-v_1} - \xiu \right) \right|_{\{\xiu,v_1\} \to \{\xit,v_3\} = \{\xi_0,v\}}~, \nn \\
&&{\cal R}_2(\xi_0,v) = \theta\left( \frac{1}{1+\sqrt{1-v}} - \xi_0 \right)~.
\label{eq:errequeerre}
\eea
The explicit expression for ${\cal R}_1(\xi_0,v)$ is, however, rather cumbersome, although this should not be 
a problem in numerical computations. For the analytic integration, we use a clever expansion of ${\cal R}_1(\xi_0,v)$
that exploits both reference systems; 
\bea
1-{\cal R}_1(\xi_0,v) &=& \theta(1-2 v) + \theta(\xi_0-1)\,  \theta(2v-1) 
+ \theta\left(\xiu-\frac{1-2v_1}{1-v_1}\right) \nn \\ &\times&
\bigg[ \theta\left(\frac{1}{1-2v_1} -\xiu \right) \, \theta\left(\frac{2-\sqrt{2}}{4} - v_1\right) \nn \\ 
&+&  \theta\left(2-4v_1 -\xiu \right) \, \theta\left(v_1-\frac{2-\sqrt{2}}{4}\right) \theta(1-2v_1) \bigg]~.
\label{eq:theta}
\eea
The first two terms in the right-hand side of~\Eq{eq:theta} contribute at large loop three-momenta 
rendering the integral defined by~\Eq{eq:remnant} finite in the UV. The next two terms provide a finite contribution.
The virtual remnant in~\Eq{eq:remnant} is also IR finite, and therefore it can be calculated with $\ep=0$.
In particular, we get
\beq
\overline \sigma^{(1)}_{\v} = \sigma^{(0)}\, a \, \frac{\pi^2}{6}
+ {\cal O}(\ep)~.
\label{eq:sigmafb}
\eeq 
The sum of all the contributions, \Eq{eq:sigma1}, \Eq{eq:sigma2} and \Eq{eq:sigmafb}, gives a total 
cross-section of ${\cal O}(\ep)$, in agreement with the result that would be obtained from the 
standard calculations in DREG.

\section{Ultraviolet renormalisation}
\label{sec:renorm}

In the previous section we have shown how to avoid the introduction of subtraction counter-terms to cancel 
soft and collinear singularities by a suitable mapping of the momenta entering virtual and real corrections. 
In any practical computation in QFT, UV divergences must also be taken into account. 
Another advantage of LTD is to enlighten the physical aspects of renormalisation.
In order to explain the proposed approach, we consider first the simplest scalar two-point function, with massless internal lines. 
There is only one external momenta, $p^{\mu}$, and the two-point function is free of IR singularities 
if the incoming momentum is not light-like. Due to the fact that the virtuality of the incoming particle is the unique 
physical scale involved in the problem, the integral vanishes if we set $p^2=0$. 
So, the non-trivial massless scalar two-point function requires $p^2 \neq 0$ and it is only UV divergent. 
Labelling the internal momenta as $q_1=\ell+p$ and $q_2=\ell$, then we have 
\beqn
L^{(1)}(p,-p) &=& \int_\ell \, \prod_{i=1}^2 G_F(q_i) =  c_\Gamma \,
\frac{\mu^{2\ep}}{\ep \, (1-2\ep)} \, \left(-p^2-\imath 0\right)^{-\ep}~,
\label{bubble}
\eeqn
as shown in the literature\footnote{For instance, see Refs. \cite{Catani:2008xa,Ellis:2007qk,Sborlini:2013jba}.}.
The LTD representation of the scalar two-point function reads
\beq
L^{(1)}(p,-p) = \sum_{i=1}^2 I_{i}~,
\eeq
with the dual integrals
\bea
\nn I_1 &=& -\int_\ell \frac{\td{q_1}}{-2\, q_1\cdot p +p^2 + \imath 0}~, \\ 
I_2 &=& -\int_\ell \frac{\td{q_2}}{2\, q_2\cdot p +p^2 - \imath 0} ~,  
\label{CorteI2BUBBLE}
\eea
where for simplicity we consider $p_0>0$ and $p^2>0$ (i.e. the incoming particle has positive energy and we work in the TL region). 
Following the discussion presented in Section~\ref{sec:triangleCUT}, we parametrise the momenta using
\beqn
p^{\mu} &=& \left(p_0, {\bf 0} \right)~, \nn
\\ q_i^{\mu} &=& p_0 \, \xi_{i,0} \, \left(1,2 \sqrt{v_i(1-v_i)} \, {\bf e}_{i,\perp}, 1-2 v_i \right)~,
\label{BUBparameter}
\eeqn
which is equivalent to settle in the rest frame of the incoming particle. With this choice, the dual integrals are rewritten as
\bea
I_1 &=& - \int \, d[\xi_{1,0}] \, d[v_1]\, \, \frac{4\xi_{1,0}}{1-2 \xi_{1,0} + \imath 0}~,  
\label{CorteI1BUBBLEv2}
\\ I_2 &=& - \int \, d[\xi_{2,0}] \, d[v_2] \, \frac{4\xi_{2,0}}{1+2 \xi_{2,0}} ~,
\label{CorteI2BUBBLEv2}
\eea
where we follow the definition of the $d$-dimensional integration measure given in \Eq{eq:measureDREG}; 
the only difference is that we must set $(2 p_0)^2=4p^2$ instead of $s_{12}$ as the normalisation scale of the system:
\beqn
d[\xi_{i,0}] = \frac{(4\pi)^{\ep-2}}{\Gamma(1-\ep)}\, \left(\frac{4 p^2}{\mu^2}\right)^{-\ep} \, \xi_{i,0}^{-2\ep}\, d\xi_{i,0}~. 
\eeqn

\begin{figure}[t]
\begin{center}
\includegraphics[width=6.5cm]{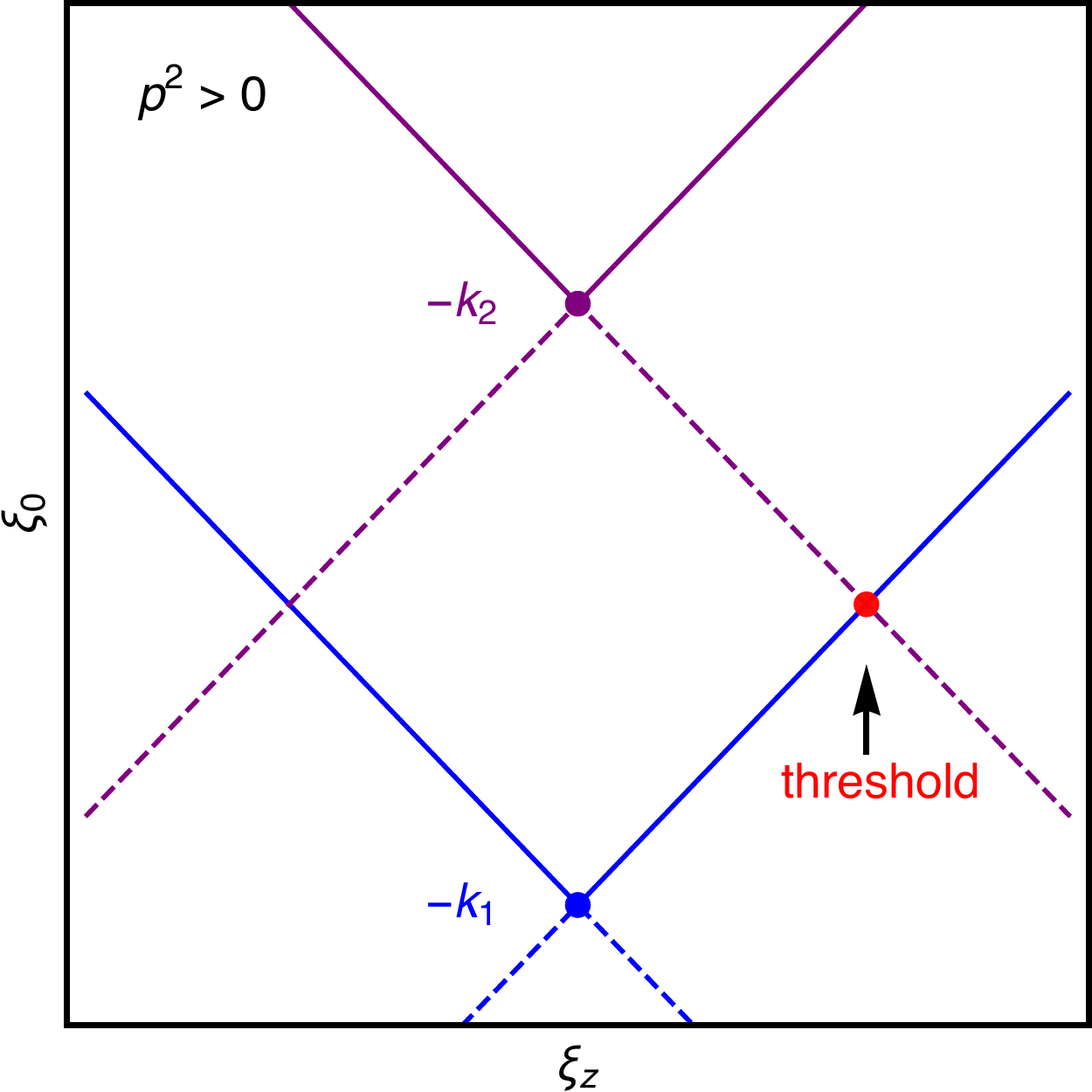} $\qquad$
\includegraphics[width=6.5cm]{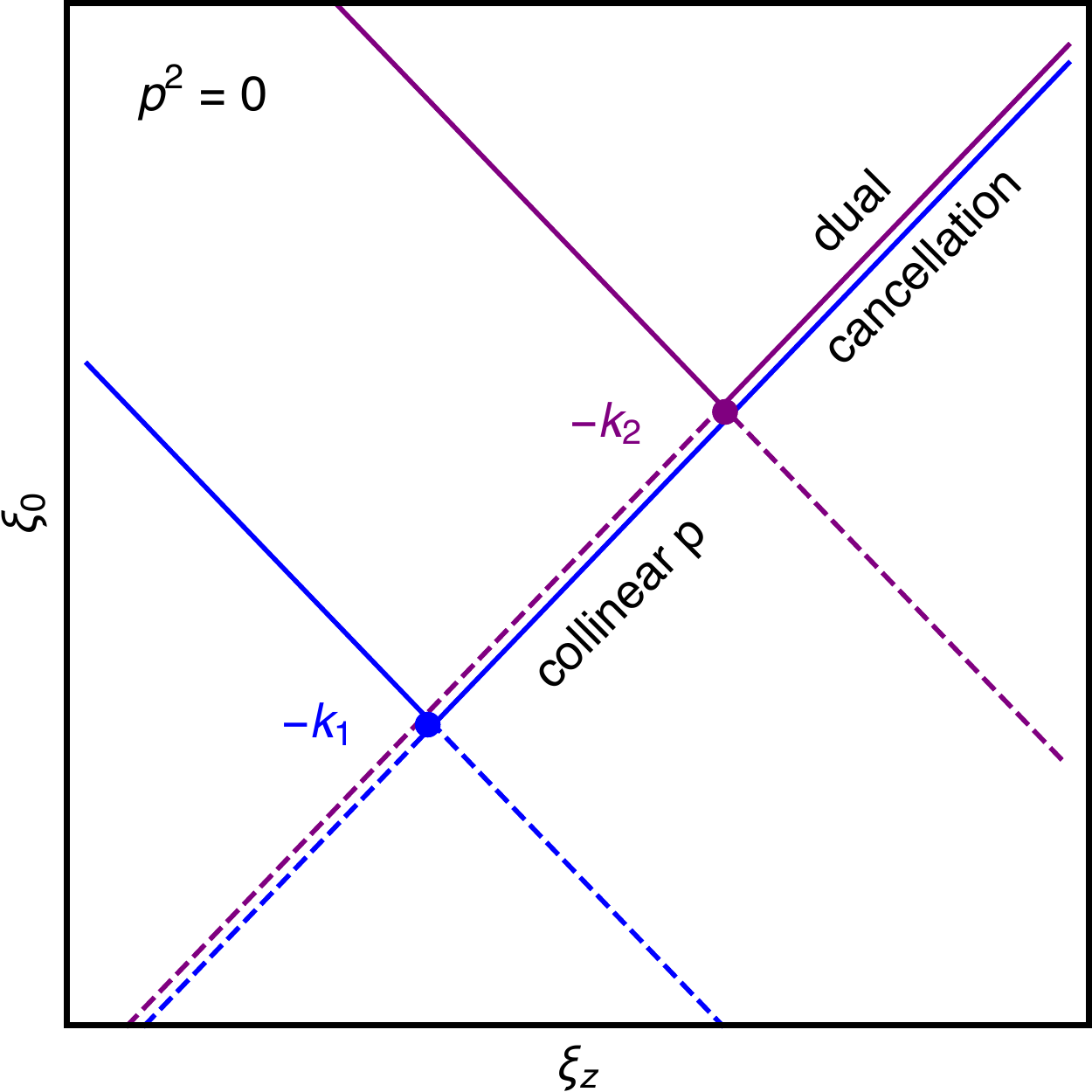}
\caption{\label{fig:bubbles} 
Light-cones of the two-point function in the loop coordinates $\ell^{\mu} = p_0\, (\xi_0, \xi_{\perp}, \xi_z)$
for a time-like, $p^2 > 0$ (left), and a light-like, $p^2 = 0$ (right), configuration. The internal momenta 
are $q_1=\ell+p$ and $q_2=\ell$. In the time-like case, a threshold singularity appears when the backward light-cone 
(dashed) of $G_F(q_2)$ intersects with the forward light-cone (solid) of $G_F(q_1)$. In the light-like case, the IR singularities 
are restricted to the compact region $\xi_{1,0} \le 1$.}
\end{center}
\end{figure}

The integration in $\xi_{i,0}$ and $v_i$ can be performed analytically, resulting in
\beqn
I_1 &=&  \frac{\cgt}{2\ep\, (1-2\ep)} \, \left(\frac{p^2}{\mu^2}\right)^{-\ep} \, e^{\imath 2 \pi \ep}~,
\label{CorteI1BUBBLEv3}
\\ I_2 &=& \frac{\cgt}{2\ep\, (1-2\ep)} \, \left(\frac{p^2}{\mu^2}\right)^{-\ep}.
\label{CorteI2BUBBLEv3}
\eeqn
The sum of both contributions gives the standard DREG result in \Eq{bubble}. 
The imaginary part of the scalar two-point function is calculable as 
\beqn
\nn \imath \, {\rm Im}\left[L^{(1)}(p,-p)\right] &=& \imath \, {\rm Im} \, I_1 = 
\frac{1}{2} \, \int_{\ell} \td{q_1} \, \td{-q_2} = \imath \, \pi \,
\int \td{q_1} \, \theta(p_{0}-q_{1,0}) \, \delta(p^2-2q_1\cdot p) \nn \\ 
&=& \imath \,  \frac{\cgt}{2\ep\, (1-2\ep)} \, \left(\frac{p^2}{\mu^2}\right)^{-\ep} \sin(2\pi\ep)~,
\label{BUBBLECutkowski1}
\eeqn
which agrees with the result directly obtained from~\Eq{bubble}. This imaginary component is associated 
with $I_1$ and it is due to the presence of a threshold singularity. This threshold  behaviour can graphically be explained 
from Fig. \ref{fig:bubbles} (left); the forward light-cone associated with $I_1$ intersects only once the backward 
region of the other light-cone, and it is the only dual integral which requires to keep explicitly 
the $+\imath 0$ prescription in \Eq{CorteI1BUBBLEv2} .

In order to build a suitable local UV counter-term of the two point-function, 
we follow the ideas presented in Ref.~\cite{Becker:2010ng} and we consider
\beqn
I^{\cnt}_{\uv} &=& \int_{\ell} \frac{1}{\left(q_{\uv}^2-\muUV^2+\imath 0 \right)^2}~, 
\label{BubbleCNTUV}
\eeqn
where $\muUV$ is an arbitrary energy scale, and $q_{\uv}=\ell+k_{\uv}$ with $k_{\uv}$ an arbitrary four-momentum. Notice that the counter-term is expressed as a Feynman integral using the customary $+\imath 0$ prescription. We should now construct the corresponding dual representation and combine it with Eqs. (\ref{CorteI1BUBBLEv3}) and (\ref{CorteI2BUBBLEv3}). The counter-term in~\Eq{BubbleCNTUV} exhibits a double pole in the complex plane of the loop energy component which
requires to apply the extended version of the LTD theorem~\cite{Bierenbaum:2012th}. 
There are two possible paths to follow: either to 
compute the residue at the relevant higher-order pole and use the Cauchy's residue theorem to perform 
the $\ell_0$ integral, or to apply integration-by-parts (IBP) identities to rewrite \Eq{BubbleCNTUV} 
in terms of a massive tadpole.

Let's start with the first approach. The residue of a function $f(z)$ at a multiple pole is given by
\beqn
{\rm Res}\left(f,z_0\right) &=& \frac{1}{(n-1)!} \, \left[\frac{\partial^{n-1}}{\partial z^{n-1}} \left((z-z_0)^n \, f(z)\right)\right]_{z=z_0} \, ,
\label{CountertermRESIDUETH2}
\eeqn
with $n$ the multiplicity of the pole at $z_0$. 
The location of the double pole of the UV propagator in~\Eq{BubbleCNTUV} 
is obtained from the on-shell condition, i.e.  
\beqn
G_F^{-1}(q_{\uv}) = q_{\uv}^2 - \mu_{\uv}^2 +\imath 0 = 0  \quad 
&\Rightarrow& \quad q_{\uv, 0}^{(\pm)} = \pm \sqrt{\qb_{\uv}^2+\muUV^2 -\imath 0} \, ,
\eeqn
where we just keep the solution $q_{\uv,0}^{(+)}$ because it lies in the lower part of the complex plane and describes a positive-energy particle. The calculation of the residue gives
\beq
{\rm Res} \left( \left(G_F(q_{\uv})\right)^2, q_{\uv,0}^{(+)}\right) = - \frac{1}{4 (q_{\uv,0}^{(+)})^3}~, 
\eeq
and
\beq
I_{\uv}^{\cnt} = \int_\ell \, \frac{\td{q_{\uv}}}{2\left( q_{\uv,0}^{(+)}\right)^2}
\label{bubbleuv}
\eeq
is a possible dual representation of the UV counter-term.

On the other hand, we can apply IBP to lower the power of the propagator, and then use LTD in its 
usual form. It is straightforward to obtain
\beqn
I^{\cnt}_{\uv} &=& \frac{1-\epsilon}{\muUV^2} \, \int_{\ell} G_F(q_{\uv})
= \frac{\ep-1}{\muUV^2} \, \int_{\ell} \, \td{q_{\uv}}~. 
\label{BUBBLECNTibpreduced}
\eeqn
Both~\Eq{bubbleuv} and~\Eq{BUBBLECNTibpreduced} are equivalent dual representations of the UV counter-term 
in~\Eq{BubbleCNTUV}. If we perform the change of variables $\xiuv = \sqrt{\qb_{\uv}^2}/p_0$, 
$m_{\uv} = \muUV /p_0$, the UV counter-term gets either of the two following dual forms:
\bea
\nn I^{\rm cnt}_{\uv} 
&=& \int \, d[\xiuv] \, d[v_{\uv}] \, \frac{2 \xiuv^2}{\left(\xiuv^2 + m_{\uv}^2\right)^{3/2}}
\\ &=&\frac{\epsilon-1}{m_{\uv}^2} \, \int \, d[\xiuv] \, d[v_{\uv}] \, 
\frac{4 \xiuv^2}{\left(\xiuv^2 + m_{\uv}^2\right)^{1/2}}~,
\label{dualuv}
\eea
which corresponds to the explicit expressions for \Eq{bubbleuv} and \Eq{BUBBLECNTibpreduced}, respectively. 
We have dropped the $+\imath 0$ prescription because it is not necessary. 
After explicit computation, both representations in \Eq{dualuv} lead to the same result: 
\beq
I^{\rm cnt}_{\rm UV} =  \frac{\Se}{(4\pi)^2}\,  
\frac{1}{\ep} \,  \left(\frac{\mu_{\uv}^2}{\mu^2}\right)^{-\ep} 
= \frac{c_{\Gamma}}{\epsilon} \left(\frac{\mu_{\uv}^2}{\mu^2}\right)^{-\ep} \, + {\cal O}(\ep^2) \, ~,  
\eeq
which successfully reproduce the single $\ep$-pole present in the scalar two-point function. 
The prefactor~\footnote{We distinguish $\Se$ from the usual ${\overline {\rm MS}}$ scheme factor  
 $\SMS = (4\pi)^{\ep} \exp(-\ep \, \gamma_E)$ or $S_{\ep} = (4\pi)^{\ep}/\Gamma(1-\ep)$ as used 
 in Ref.~\cite{Bolzoni:2010bt}. At NLO all these definitions lead to the same expressions. 
 At NNLO, they lead to slightly different bookkeeping of the IR and UV poles at intermediate steps, 
 but physical cross-sections of infrared safe observables are the same. } is defined as 
 $\Se = (4\pi)^{\ep} \Gamma(1+\ep)$.

The next step consists in combining a dual representation of the UV counter-term with the dual integrals $I_1$ and $I_2$. 
So, we define the renormalised scalar two-point function as
\beqn
\nn L^{(1,\r)}(p,-p) &=& L^{(1)}(p,-p) - I_{\uv}^{\cnt} \\ 
&=& - 4 \int \, d[\xi] \, d[v] \, \left[\frac{\xi}{1-2 \xi+\imath 0}+\frac{\xi}{1+2 \xi}
+\frac{\xi^2}{2 \left(\xi^2+m_\uv^2 \right)^{3/2}} \right]~,
\label{Bubbleregular1}
\eeqn
with $\xi=\xi_{i,0}=\xi_{\uv}$ and $v=v_i=v_{\uv}$, that verifies
\beqn
L^{(1,\r)}(p,-p)  &=& \frac{1}{(4\pi)^2} \left[-\ln{-\frac{p^2}{\muUV^2} - \imath 0} + 2 \right] + {\cal O}(\ep)~,
\label{Bubbleregular2}
\eeqn
which is free of $\epsilon$-poles. The integrand of \Eq{Bubbleregular1} is integrable in the limit $\ep \to 0$, 
and the computation can be fully performed with $\ep=0$.  So, we succeeded in finding a purely 
four-dimensional representation of the renormalised two-point function. 
Consistently, the integration and the limit $\ep\to 0$ commute.

\subsection{Scaleless two-point function}
\label{ssec:masslessbubbles}

As mentioned before, if we consider $p^2=0$ then $L^{(1)}(p,-p)=0$ since it does not contain any scale. From a physical point of view, this implies that self-energy corrections to on-shell massless particles are zero. However, a vanishing integral in DREG does not imply a vanishing integrand. 
This is particularly an issue in LTD where the aim is to cancel singularities locally, and therefore it is relevant to properly characterise the IR and 
the UV behaviour separately. So, let's consider the LTD representation of the massless two-point function in the light-like case, i.e. $p^2=0$. We start from \Eq{CorteI2BUBBLE}, and use the parametrisation
\beq
p^\mu = p_0 \, (1,{\bf 0}_{\perp},1) \, ,
\eeq
for the incoming particle, whilst $q_i^{\mu}$ is given by \Eq{BUBparameter}. Then, the dual contributions become
\bea
\nn I_1 &=& \int \, d[\xi_{1,0}] \, d[v_1] \, v_1^{-1} ~,  
\\ I_2 &=& - \int \, d[\xi_{2,0}] \, d[v_2] \, v_2^{-1}~,
\label{eq:BUBBLEMASSLESScortadas}
\eea
whose sum is equal to zero because they have opposite sign. In the context of DREG, 
the scaleless scalar two-point function develops both IR and UV divergences that cancel each other 
because the parameter $\ep_{\ir}$ regularising the IR singularities is identified with $\ep_{\uv}$, which regulates the UV divergences. 
The important fact is that we can exploit LTD to separate them, proceeding analogously as we did in Section~\ref{sec:triangleCUT}. Analysing the integration domain of the dual contributions, and the relative position of the light-cone as shown in Fig.~\ref{fig:bubbles}  (right),
we realise that the IR singularity is associated with $I_1$, because its forward light-cone overlaps with the backward 
light-cone of $G_F(q_2)$ in the region $\xi_{1,0}\le 1$. Thus, we define
\beq
\left. L^{(1)}_{\ir}(p,-p)\right|_{p^2=0} = I_1 (\xi_{1,0}\le 1) = 
- \frac{\cgt}{\ep(1-2\ep)} \, \left(\frac{4 p_0^2}{\mu^2}\right)^{-\ep}\, \frac{\sin(2\pi\ep)}{2\pi\ep}~,
\label{eq:LRmasslessbubble}
\eeq
which contains a single $\ep$-pole. Again, as we found for the massless three-point function, the IR singularities are confined in a compact region of the loop three-momentum space. Outside this region, the remnant is given by 
\beq
\left. L^{(1)}_{\uv}(p,-p)\right|_{p^2=0} = I_1 (\xi_{1,0} > 1) + I_2 = 
\frac{\cgt}{\ep(1-2\ep)} \, \left(\frac{4 p_0^2}{\mu^2}\right)^{-\ep}\, \frac{\sin(2\pi\ep)}{2\pi\ep}~,
\label{eq:LUVmasslessbubble}
\eeq
which is renormalised with the UV counter-term defined in~\Eq{bubbleuv}. It is worth noting that LTD leads naturally to this separation of IR/UV regions, which is crucial to achieve a local cancellation of singularities in the computation of physical observables at higher-orders.

\subsection{Renormalisation of scattering amplitudes and physical interpretation}
\label{ssec:generalUV}

In general, the UV counter-terms of scattering amplitudes are derived by expanding the internal propagators 
around the UV propagator~\cite{Becker:2010ng}. For a single propagator:
\bea
\frac{1}{q_i^2-m_i^2+\imath 0} &=&
\frac{1}{q_{\uv}^2-\mu_{\uv}^2+\imath 0} \\ &\times&
\bigg[1 - \frac{2q_{\uv}\cdot k_{i,\uv}  + k_{i,\uv}^2-m_i^2+\mu_{\uv}^2}{q_{\uv}^2-\mu_{\uv}^2+\imath 0}
+ \frac{(2q_{\uv}\cdot k_{i,\uv})^2 }{(q_{\uv}^2-\mu_{\uv}^2+\imath 0)^2} \bigg] + {\cal O} \left( (q_{\uv}^2)^{-5/2}\right)~,  \nn
\eea
with $k_{i,\uv} = q_i-q_{\uv}$, and similarly with numerators. In order to improve the convergence in numerical implementations, 
the authors of Ref.~\cite{Becker:2012aqa} propose to expand even to higher-powers of $G_F(q_{\uv})^{-1}$. 

Once the desired UV expansion is obtained, we shall derive the corresponding dual representation 
following the procedure described before. To obtain a suitable LTD representation of the UV counter-terms, 
it is necessary to deal with multiple-poles and non-trivial numerators depending on $q_{\uv}$.
The following identities are enough to construct most of the UV counter-terms:
\bea
&& \int_\ell \, \left( G_F(q_{\uv}) \right)^n = 
\frac{(-1)^n \, (2n-2)!}{((n-1)!)^2}
\int_\ell \, \frac{\td{q_{\uv}}}{\left( 2 q_{\uv,0}^{(+)}\right)^{2n-2}}~, \nn \\ 
&& \int_\ell q_{\uv,0} \, \left( G_F(q_{\uv}) \right)^n = 0~, \nn \\
&& \int_\ell \, q_{\uv,0}^2 \, \left( G_F(q_{\uv}) \right)^n = 
\frac{(-1)^{n-1} \, (2n-4)!}{2\, (n-1)!\, (n-2)!}
\int_\ell \, \frac{\td{q_{\uv}}}{\left( 2 q_{\uv,0}^{(+)}\right)^{2n-4}}~. 
\eea
These expressions have been obtained from \Eq{CountertermRESIDUETH2}. 
Alternatively, we can apply IBP to obtain equivalent dual representations. 
Explicit examples will be presented in Section~\ref{sec:gammatoqqbar}.
Only the genuine UV singularities of the original scattering amplitudes 
need to be subtracted with this procedure. The spurious UV singularities 
of the individual dual integrals are cancelled in the sum of all the dual contributions. 

\begin{figure}[t]
\begin{center}
\includegraphics[width=8cm]{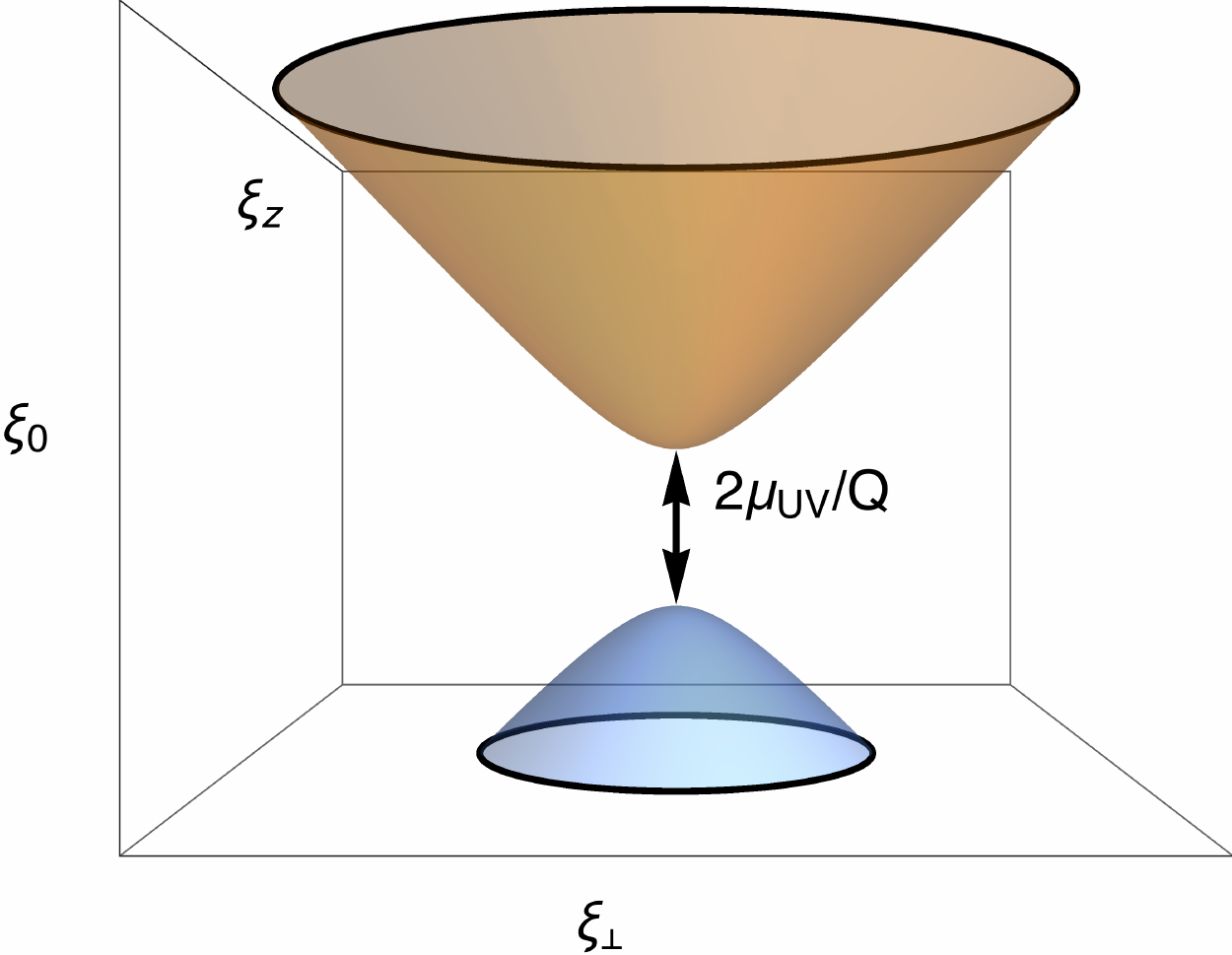}
\caption{On-shell hyperboloids of the UV counter-term.  The forward and backward on-shell hyperboloids
are separated by a distance $2\mu_{\uv}/Q$, where $Q$ is the characteristic hard scale. 
\label{fig:carteseanUV}}
\end{center}
\end{figure}

To conclude this section, let us remind about the physical interpretation of the 
arbitrary energy scale  $\muUV$ introduced in the renormalisation procedure~\cite{Hernandez-Pinto:2015ysa}.
This arbitrary scale can be interpreted as a renormalisation scale, since the UV counter-term affects the behaviour of the integrand only in the high-energy region. In fact, as seen in Fig.~\ref{fig:carteseanUV}, its dual representation only contributes for loop energies larger than $k_{\uv,0}+\mu_{\uv}$, although it is unconstrained in the loop three-momentum. So, we should choose $\mu_{\uv} \geq {\cal O}(Q)$, with $Q$ the physical hard scale that determines the size of the compact region in the loop momentum space where the IR and threshold singularities are located. 
However, the intersection of the integration domains of the dual integrands gives rise to singularities, 
thus we could improve the UV cancellations if we avoid those intersections.
In other terms, we must choose $\mu_{\uv}$ in such a way that the on-shell hyperboloids of the UV propagator do not intersect with any 
of the on-shell hyperboloids of the original integral. Since the UV forward and backward on-shell hyperboloids are separated by a distance 
$2 \mu_{\uv}$, it is possible to give a physical motivation for an optimal minimal choice of $\mu_{\uv}$ and $k_{\uv}$. 
If we take $\mu_{\uv} = Q/2$, and $k_\uv$ in the centre of the physical compact region, these conditions are fulfilled in a minimal way, i.e. we naturally avoid intersections with the physical on-shell hyperboloids.

\section{NLO corrections to $\gamma^* \to q \bar q (g)$}
\label{sec:gammatoqqbar}

In this section, we discuss in detail the computation of NLO QCD corrections to the total cross-section for the 
process $\gamma^* \to q\bar q (g)$ by using the LTD approach. We emphasise that it constitutes the first realistic 
physical application of this method, as already anticipated in Ref. \cite{Sborlini:2016fcj}.

\begin{figure}[t]
\begin{center}
\includegraphics[width=0.95\textwidth]{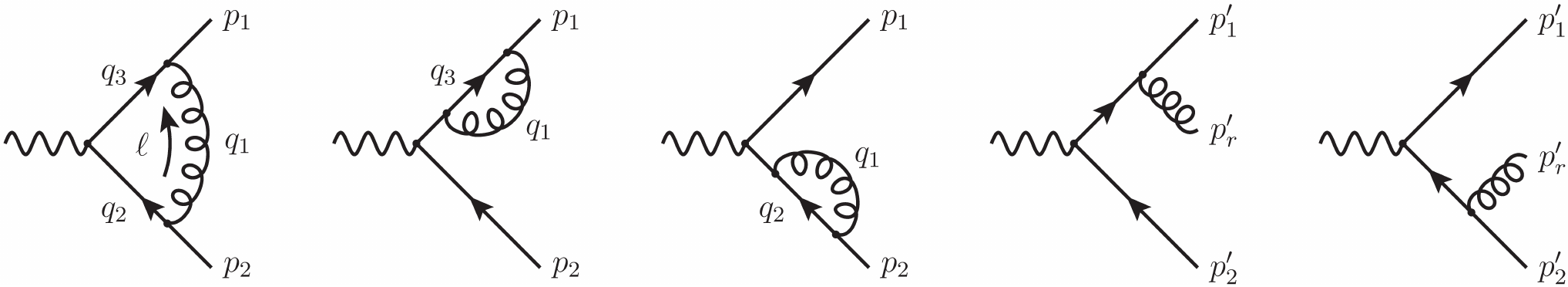}
\caption{Momentum configuration and Feynman diagrams associated with the process $\gamma^* \to q \bar{q} (g)$, up to ${\cal O}(\alpha\, \as)$. 
Notice that we also consider self-energy corrections to the on-shell outgoing particles, even if their total contribution is zero.
\label{fig:PHOTONdecay}}
\end{center}
\end{figure}

The computation is done in the context of QCD$+$QED with massless quarks, up to ${\cal O}(\alpha \, \as)$. The requested  Feynman diagrams are shown in Fig. \ref{fig:PHOTONdecay}. Starting at the LO, we have
\beq
|\M{0}_\qq|^2 = 2 \, C_A \, (e\, e_q)^2 \, s_{12} \, (1-\ep) \ , 
\eeq
for $\gamma^*(p_{12}) \to q(p_1) + \bar q(p_2)$ with $p_{12}=p_1+p_2$ and $p_{12}^2=s_{12}>0$, where $e$ and $e_q$ 
denote the electromagnetic coupling and the quark electric charge, respectively\footnote{As usual, the squared matrix elements 
are averaged over the number of spin degrees of freedom of the incoming particles, which is taken to be $2(1-\ep)$ for the photon.
In any case, we normalise the NLO results by the LO contribution.}. 
The corresponding Born level total cross-section is given by
\beq
\sigma^{(0)} = \frac{1}{2s_{12}} \int d\Phi_{1\to 2} \,  |\M{0}_\qq|^2 =  \frac{1}{2} \, \alpha \, e_q^2 \, C_A +{\cal O}(\ep)\, ,
\eeq
where the two-body phase-space factor is shown in \Eq{Medida12INTEGRADA} (Appendix~\ref{app:phasespace}). 

Let's centre in the NLO contributions. The real corrections from the radiative process $\gamma^*(p_{12})\to q(p_1') + \bar{q}(p_2') + g(p_r')$ are 
\bea
\sigma_{\rm R}^{(1)} &=& \sigma^{(0)}  \, 
\frac{(4\pi)^{\ep-2}}{\Gamma(1-\ep)} \, \gs^2 \, C_F \, 
\left(\frac{s_{12}}{\mu^2}\right)^{-\ep} \,  
\int_0^{1} dy_{1r}'  \, \int_0^{1-y_{1r}'} dy_{2r}' \, (y_{1r}' \, y_{2r}' \,y_{12}' )^{-\ep}\, 
\nn \\ &\times& \left[ 4\left( \frac{y_{12}'}{y_{1r}'\, y_{2r}'} - \ep \right)
+ 2(1-\ep) \left( \frac{y_{2r}'}{y_{1r}'} + \frac{y_{1r}'}{y_{2r}'} \right)  \right]
\nn \\ 
&=& \sigma^{(0)} \, c_{\Gamma} \, \gs^2 \, C_F \,  \left(\frac{s_{12}}{\mu^2}\right)^{-\ep} \,  
\frac{4 (2 - 2\ep + \ep^2) \, \Gamma(2-2\ep)}
{\ep^2\, \Gamma(3-3\ep) \, \Gamma(1+\ep)}~, 
\label{eq:realtotal}
\eea
where the virtuality of the photon is $s_{12}$, the final-state momenta are denoted primed to distinguish them 
from the Born level kinematics, and $y_{ij}'=2 \, p_i' \cdot p_j' / s_{12}$. To obtain the expressions in \Eq{eq:realtotal} we use the expansion of the three-body phase-space shown in \Eq{Medida13ORIGINALrelativa}. If we expand this result, we find
\beq
\sigma_{\rm R}^{(1)} = \sigma^{(0)} \, c_{\Gamma} \, \gs^2 \, C_F \,  \left(\frac{s_{12}}{\mu^2}\right)^{-\ep} \, 
\left[\frac{4}{\ep^2}+\frac{6}{\ep}+ 19 - 2\pi^2 \, + {\cal O}(\ep)\right]~, 
\label{eq:realtotalseries}
\eeq
which contains both double and single $\epsilon$-poles, associated with soft and collinear singularities, respectively. On the other hand, the virtual contribution is generated by the interference of the one-loop vertex correction 
with the Born amplitude, which is given by 
\bea
\la \M{0}_\qq|\M{1}_\qq \ra &=& 
- |\M{0}_\qq|^2 \,  \frac{\gs^2 \,C_F}{4 \, s_{12} \, (1-\ep)} \, 
\int_\ell \left( \prod_{i=1}^3 G_F(q_i) \right) \nn \\ &\times&
{\rm Tr} \left( \bp_1 \gamma^{\nu_1} \bq_3 \gamma^{\mu_1} \bq_2 \gamma^{\nu_2} 
\bp_2 \gamma^{\mu_2} \right) \, d_{\nu_1 \nu_2}(q_1) \, d_{\mu_1 \mu_2}(p_{12})~,
\label{eq:vertexqqg}
\eea
with $d_{\nu_1 \nu_2}(p)$ the gluon (or photon) polarisation tensor. In order to obtain this expression, the internal momenta 
are defined as $q_1=\ell+p_1$, $q_2=\ell+p_{12}$ and $q_3=\ell$, and are considered as in the scalar case discussed in 
Section~\ref{sec:triangleREAL}. In the Feynman gauge, the expression 
in \Eq{eq:vertexqqg} takes the form
\bea
\la \M{0}_\qq|\M{1}_\qq \ra &=& 
\gs^2 \, C_F \, |\M{0}_\qq|^2 \, \frac{4}{s_{12}}
\int_\ell \left( \prod_{i=1}^3 G_F(q_i) \right) \nn \\ &\times&
\left[ (2+\ep) (q_2\cdot p_1)(q_3\cdot p_2) - \ep \left( (q_2\cdot p_2)(q_3\cdot p_1) 
+ \frac{s_{12}}{2}\, (q_2\cdot q_3) \right) \right]~,
\label{eq:vertexqqg2}
\eea
and leads to the following dual contributions
\bea
\la \M{0}_\qq|\M{1}_{\qq,1} \ra &=& - 2 \, 
\gs^2 \, C_F \, |\M{0}_\qq|^2 \, \int d[\xi_{1,0}] \, d[v_1] \, 
\left(\xi_{1,0}^{-1} \, v_1^{-1} + 1 \right) \left((1-v_1)^{-1} - \xi_{1,0}\right)~, \nn \\
\la \M{0}_\qq|\M{1}_{\qq,2} \ra &=& - 2 \, 
\gs^2 \,  C_F \, |\M{0}_\qq|^2 \, \int d[\xi_{2,0}] \, d[v_2] \, 
\frac{\xi_{2,0}}{1-\xi_{2,0}+\imath 0} \, 
\left( v_2 \, ( (1-v_2)^{-1} - \xi_{2,0}) - \ep \right)~, \nn \\
\la \M{0}_\qq|\M{1}_{\qq,3} \ra &=& - 2 \, 
\gs^2 \,  C_F \, |\M{0}_\qq|^2 \, \int d[\xi_{3,0}] \, d[v_3] \, 
\frac{\xi_{3,0}}{1+\xi_{3,0}} \, 
\left( (1-v_3) \, ( v_3^{-1} + \xi_{3,0} ) - \ep \right)~,
\label{eq:dualsqqg}
\eea
with $|\M{1}_\qq \ra = \sum |\M{1}_{\qq,i} \ra$. The individual dual integrals in~\Eq{eq:dualsqqg} 
contain up to quadratic UV divergences, although the naive power counting in the original loop integral 
in~\Eq{eq:vertexqqg} leads to a logarithmic behaviour in the UV.
The quadratic UV divergences cancel in the sum of all the dual integrals, 
while the linear UV divergences disappear after integration over the polar angle.
Applying the change of variables from~\Eq{eq:changexi10} in the first dual contribution, 
however, produces the cancellation of the linear UV divergences also at the integrand level. 
As expected, only the logarithmic UV divergences remain in the sum of the dual contributions. 
Performing the explicit integration over the loop variables leads to 
\bea
\la \M{0}_\qq|\M{1}_{\qq,1} \ra &=& 0~,  \nn \\
\la \M{0}_\qq|\M{1}_{\qq,2} \ra &=& |\M{0}_\qq|^2 \, \cgt \,
\gs^2 \,  C_F \, \left(\frac{s_{12}}{\mu^2}\right)^{-\ep} \, 
\frac{1}{\ep^2} \left( \frac{\ep}{2} - \frac{1}{1-2\ep} \right) 
 \, e^{\imath 2\pi\ep}~, \nn \\
\la \M{0}_\qq|\M{1}_{\qq,3} \ra &=& |\M{0}_\qq|^2 \, \cgt \,
\gs^2 \,  C_F \, \left(\frac{s_{12}}{\mu^2}\right)^{-\ep}
\, \frac{1}{\ep^2} \left( \frac{\ep}{2} - \frac{1}{1-2\ep} \right) ~.
\eea
Putting together the three dual terms, we obtain
\beqn
\nn \sigma_{\rm V}^{(1)}
&=& \sigma^{(0)} \, c_{\Gamma} \, \gs^2 \, C_F \,  \left(\frac{s_{12}}{\mu^2}\right)^{-\ep} \,  
\frac{2}{\ep^2} \left( \ep - \frac{2}{1-2\ep} \right) \cos(\pi \ep)~
\\ &=& \sigma^{(0)} \, c_{\Gamma} \, \gs^2 \, C_F \,  \left(\frac{s_{12}}{\mu^2}\right)^{-\ep} \, 
\left[-\frac{4}{\ep^2}-\frac{6}{\ep} - 16 + 2\pi^2 \, + {\cal O}(\ep)\right]  \, , 
\label{eq:virtualtotal}
\eeqn
i.e., we recover the virtual contribution to the total cross-section at NLO. 
Notice that it was unnecessary to introduce any tensor reduction; Gram determinants 
are naturally avoided in LTD, and therefore also are the spurious singularities that the tensor reduction 
introduces leading to numerical instabilities in the integration over the phase-space.
Finally, if we sum the contributions from \Eq{eq:realtotal} and \Eq{eq:virtualtotal}, we obtain
\beq
\sigma = \sigma^{(0)} \left( 1 +  3 \, C_F \, \frac{\as}{4\pi}  
+ {\cal O}(\as^2) \right)~,
\label{eq:totalNLO}
\eeq
which agrees with the well-known result available in the literature. 
The $\epsilon$-poles cancel between real and virtual contributions 
(as expected from the KLN theorem), so we can safely take the limit $\ep \to 0$ \emph{after} integration. 

It is the purpose of this section to show that $\ep \to 0$ can be considered also \emph{before} integration, once a proper combination of real and virtual terms is done. In the context of LTD, we shall also consider carefully the contributions introduced by self-energy diagrams. On-shell massless quarks do not introduce further corrections to the total cross-section in~\Eq{eq:totalNLO} due to the renormalisation of the wave function because IR and UV divergences are treated equally in DREG. In the on-shell scheme
the wave function renormalisation constant contains both IR and UV divergences, but they cancel each other, which justifies the exclusion of the corresponding Feynman diagrams when carrying out the computation within the traditional approach. 

In order to build a complete LTD representation of the virtual contributions, it is required to include the renormalised self-energy corrections to the external particles and properly disentangle IR/UV singularities at integrand level. This step is crucial to achieve a dual representation with a fully local cancellation of singularities, so it can be integrated in four-dimensions. The quark and antiquark self-energies at one-loop are given by
\bea
-\imath \, \Sigma(p_1) &=& \imath \, \gs^2 \, C_F \,  \int_\ell \left( \prod_{i=1,3} G_F(q_i) \right)
\gamma^\mu (-\bq_3+m)  \gamma^\nu \,  d_{\mu\nu} (q_1)~,  \nn \\
-\imath \, \Sigma(-p_2) &=& \imath \, \gs^2 \, C_F \, \int_\ell \left( \prod_{i=1,2} G_F(q_i) \right) 
\gamma^\mu (-\bq_2+m)  \gamma^\nu \,  d_{\mu\nu}(q_1)~,
\label{eq:selfenergies}
\eea
with $\Sigma(p_i) = \Sigma_2 \, \bp_i -  \Sigma_1 \, m$. 
In these expressions, we keep the same internal momenta $q_i$ that were used to define the vertex corrections in \Eq{eq:vertexqqg}. 
From the usual renormalisation procedure, the self-energy contribution is related with the renormalisation factor $Z_2 = 1+\Delta Z_2$. 
Applying on-shell renormalisation conditions to the quark and antiquark self-energies
in the Feynman gauge, we obtain the following contributions
\bea
\la \M{0}_{\qq} | \Sigma(p_1) \ra &=&  - 2(1-\ep) \, \gs^2 \, C_F \, |\M{0}_{\qq}|^2 \,
\int_\ell \left( \prod_{i=1,3} G_F(q_i) \right)  \, 
\left( 1+ \frac{q_3\cdot p_2}{p_1\cdot p_2} \right)~, 
\label{eq:renormalizationconstants1} \\ 
\la \M{0}_{\qq} | \Sigma(p_2) \ra  &=&  - 2(1-\ep)\, \gs^2 \, C_F \, |\M{0}_{\qq}|^2 \,
\int_\ell \left( \prod_{i=1,2} G_F(q_i) \right)  \, 
\left( 1 - \frac{q_2\cdot p_1}{p_1\cdot p_2}  \right)~.
\label{eq:renormalizationconstants2}
\eea
Formally, these contributions vanish
in DREG, but they feature a non-trivial IR and UV behaviour at the integrand level 
that we need to make explicit in order to have a local cancellation 
of all the singularities with those present in the real corrections. Applying LTD to the loop integrals 
given in Eqs. (\ref{eq:renormalizationconstants1}) and (\ref{eq:renormalizationconstants2}), 
the corresponding dual representations are
\bea
\la \M{0}_{\qq} | \Sigma(p_1) \ra &=& -2 (1-\ep) \, \gs^2 \, C_F \, |\M{0}_{\qq}|^2 \,\bigg[ 
\int d[\xiu] \, d[v_1] \, v_1^{-1} \, (1-v_1) \, \xiu \nn \\ &-&  
\int d[\xit] \, d[v_3] \, v_3^{-1} \, (1+(1-v_3) \, \xit)  \bigg]~, \\
\la \M{0}_{\qq} | \Sigma(p_2) \ra &=& -2 (1-\ep) \, \gs^2 \, C_F \,  |\M{0}_{\qq}|^2 \,\bigg[ 
\int d[\xiu] \, d[v_1] \, v_1\, (1-v_1)^{-1}\,  \xiu \nn \\ &+&  
\int d[\xid] \, d[v_2] \, (1-v_2)^{-1} \, (1-v_2\, \xid)  \bigg]~,
\label{eq:dualrenormalization}
\eea
where we also kept explicitly the integration variables associated with each cut.

The UV divergences of the wave function
cancel exactly the UV divergences of the vertex corrections, 
because conserved currents or partially conserved currents, as the
vector and axial ones, do not get renormalised. To achieve a local 
cancellation of the UV divergences, it is relevant to note that the 
vertex corrections diverge logarithmically in the UV, while the 
expressions in~Eqs. (\ref{eq:renormalizationconstants1}) and (\ref{eq:renormalizationconstants2}) behave linearly in 
the UV. However, the linear UV divergence cancels upon angular integration. 
Therefore, a subtraction UV counter-term is needed to cancel locally also  
the linear singularities. Assuming $k_{\uv}=0$ (namely $q_{\uv} = \ell$) 
and following the discussion of Section~\ref{sec:renorm}, we define
\bea
\nn \la \M{0}_{\qq} | \Sigma_{\uv}(p_1) \ra &=& -2\, (1-\ep) \, \gs^2 \,  C_F \,  |\M{0}_{\qq}|^2 \,
\int_\ell \, \left(G_F(q_{\uv})\right)^2 \, 
\left(1 + \frac{q_{\uv} \cdot p_2}{p_1\cdot p_2} \right) 
\\ &\times& \left[ 1 -  G_F(q_{\uv}) (2\, q_{\uv} \cdot p_1 + \mu^2_{\uv}) \right]~, 
\eea
whose dual representation is given by
\bea
\la \M{0}_{\qq} | \Sigma_{\uv}(p_1) \ra &=& -2 (1-\ep) \, \gs^2 \,  C_F \,  |\M{0}_{\qq}|^2 \,
\nn \\ && \qquad \times \int_\ell 
\frac{\td{q_{\uv}}}{2\left(q_{\uv,0}^{(+)}\right)^2} 
\bigg[ \bigg(1 - \frac{\qb_{\uv}\cdot \pb_2}{p_1\cdot p_2} \bigg) \, 
\bigg( 1 - 
\frac{3\, (2\, \qb_{\uv}\cdot \pb_1 - \mu^2_{\uv} )}
{4 \left(q_{\uv,0}^{(+)}\right)^2} \bigg) - \frac{1}{4} \bigg]
\nn \\
&=& - 2 (1-\ep) \, \gs^2 \,  C_F \, |\M{0}_{\qq}|^2 \, \int d[\xiuv] \, d[v_{\uv}] \, 
\frac{\xiuv^2}{(\xiuv^2 + m^2_{\uv})^{3/2}} \nn \\ && \qquad \times 
\left[ \bigg(2 + \xiuv\, (1-2v_{\uv})\bigg) \left( 1 - 
\frac{3 \, \left( 2 \, \xiuv \, (1-2v_{\uv}) - m^2_{\uv}\right)}
{4\, (\xiuv^2+m_{\uv}^2)}\right)
- \frac {1}{2}\right]~, \nn \\
\label{eq:DZetaUV}
\eea
with $m_{\uv} = 2 \mu_{\uv}/\sqrt{s_{12}}$. Notice that the term proportional to $(1-2v_{\uv})$ integrates to zero and cancels
the linear UV singularity. The integration of \Eq{eq:DZetaUV} leads to
\beq
\la \M{0}_{\qq} | \Sigma_{\uv}(p_1) \ra = -\Se \, \aas \, C_F \, |\M{0}_{\qq}|^2 \,
\left( \frac{\mu_{\uv}^2}{\mu^2}\right)^{-\ep} 
\frac{1}{\ep}  + {\cal O}(\ep)~.
\label{eq:Z2divergence}
\eeq
Consequently, $\la \M{0}_{\qq} | \Sigma(p_1) - \Sigma_{\uv}(p_1)\ra$ develops IR singularities only.    
The same counter-term cancels the UV divergence of the antiquark leg, i.e. 
$\la \M{0}_{\qq} | \Sigma_{\uv}(p_2) \ra = \la \M{0}_{\qq} | \Sigma_{\uv}(p_1) \ra$. 

Similarly, it is necessary to remove the UV divergences from the vertex correction included in $|\M{1}_\qq \ra$. 
In fact, working in the Feynman gauge, we use the following counter-term
\bea
\la \M{0}_\qq|\M{1}_{\qq,\uv} \ra &=& \, 
\gs^2 \, C_F \, |\M{0}_\qq|^2 \, 
\int d[\xiuv] \, d[v_{\uv}] \, \nn \\ &\times& 
\frac{\xiuv^2 \left( 4 (1 - 3v_{\uv} (1-v_{\uv})-\ep) \, \xiuv^2
+(7-4\ep) \, m_{\uv}^2 \right)}
{(\xiuv^2+m_{\uv}^2)^{5/2}}~, 
\eea
where the subleading terms have been fixed such that only the pole 
is subtracted, i.e.
\bea
\la \M{0}_\qq|\M{1}_{\qq,\uv} \ra &=& \Se \, \aas \, C_F \, |\M{0}_{\qq}|^2 \,
\left( \frac{\mu_{\uv}^2}{\mu^2}\right)^{-\ep} 
\frac{1}{\ep} \, .
\label{eq:M1divergence}
\eea
The crucial observation here is that \Eq{eq:Z2divergence} and \Eq{eq:M1divergence} share the same divergent structure, 
which allows to cancel completely the UV divergences.

With all these ingredients, we are able to define a four-dimensional 
representation of the total cross-section at NLO, with a local cancellation of the
IR divergences of the loop and the real corrections. 
In first place, the UV renormalised virtual cross-section is given by
\beq
\sigma_{\v}^{(1,\r)} = \frac{1}{2s_{12}} \, \int d\Phi_{1\to 2} \, 2 {\rm Re} 
\la \M{0}_\qq| \M{1,\r}_\qq \ra~,
\label{eq:virtualqqg}
\eeq
with $\M{1,\r}_\qq = \M{1}_\qq - \M{1}_{\qq,\uv}$, which also includes the self-energy corrections for simplicity . 
The renormalised virtual cross-section $\sigma_{\v}^{(1,\r)}$ 
contains only IR singularities at the integrand level, and the UV counter-terms involve a non-trivial integrand level cancellation 
of UV singularities that must be taken into account in order to find a proper four-dimensional representation of the 
total cross-section. For this reason, we start by splitting $\sigma_{\v}^{(1,\r)}$ according to
\beq
\sigma_{\v}^{(1,\r)} = \sigma_{\v}^{(1)} - \sigma_{\v}^{(1,\uv)} \, ,
\label{eq:UVsplitting}
\eeq
where we define
\bea
\sigma_{\v}^{(1)} &=& \frac{1}{2s_{12}} \, \int d\Phi_{1\to 2} \, 2 {\rm Re} 
\la \M{0}_\qq|\M{1}_\qq \ra~,
\label{eq:virtualqqgUVpart}
\\ \sigma_{\v}^{(1,\uv)} &=& \frac{1}{2s_{12}} \, \int d\Phi_{1\to 2} \, 2 {\rm Re} 
\la \M{0}_\qq|{\cal M}_{\qq,\uv}^{(1)} \ra~,
\label{eq:virtualqqgUVcouterterm}
\eea
as the original virtual terms (including self-energies) and the UV counter-terms, respectively. From \Eq{eq:virtualqqgUVpart}, we collect all the 
dual terms arising when either of the internal momenta $q_1$ or $q_2$ are set on-shell, 
and restrict the loop integration by the dual mapping conditions defined 
in~\Eq{eq:mappingconditions1} and~\Eq{eq:mappingconditions2}, respectively. This leads us to define the virtual dual contributions to the cross-section as
\bea
\widetilde \sigma_{\v,1}^{(1)} &=& - \sigma^{(0)} \, \gs^2 \, C_F \,  
\int d[\xi_{1,0}] \, d[v_1] \, {\cal R}_1(\xiu,v_1) \, \bigg[ 
4 \left(\xi_{1,0}^{-1} \, v_1^{-1} + 1\right) \left((1-v_1)^{-1} - \xi_{1,0}\right)~ \nn \\
&+& 2 (1-\ep) \, \xiu \, \left( v_1^{-1}(1-v_1) + v_1 (1-v_1)^{-1} \right) \bigg]~, \nn \\
\widetilde \sigma_{\v,2}^{(1)} &=& - \sigma^{(0)} \, \gs^2 \, C_F \,  
\int d[\xi_{2,0}] \, d[v_2] \, {\cal R}_2(\xid,v_2) \, \bigg[ 
 \frac{4 \xid}{1-\xid} \left(v_2 ((1-v_2)^{-1}-\xid)-\ep \right) \nn \\ 
&+& 2 (1-\ep) \, (1-v_2)^{-1} (1-v_2 \, \xid) \bigg]~,
\label{eq:dualvirtualcross}
\eea
together with the dual remnants
\bea
\bar \sigma_{\v,1}^{(1)} &=& - \sigma^{(0)} \, \gs^2 \, C_F \,  
\int d[\xi_{1,0}] \, d[v_1] \, (1-{\cal R}_1(\xiu,v_1)) \, \bigg[ 
4 \left(\xi_{1,0}^{-1} \, v_1^{-1} + 1\right) \left((1-v_1)^{-1} - \xi_{1,0}\right)~ \nn \\
&+& 2 (1-\ep) \, \xiu \, \left( v_1^{-1}(1-v_1) + v_1 (1-v_1)^{-1} \right) \bigg]~, \nn \\
\bar \sigma_{\v,2}^{(1)} &=& - \sigma^{(0)} \, \gs^2 \, C_F \,  
\int d[\xi_{2,0}] \, d[v_2] \, (1-{\cal R}_2(\xid,v_2)) \, \bigg[ 
 4\, \xid \, \left(v_2 ((1-v_2)^{-1}-\xid)-\ep \right)  \nn \\
&& \times \left( \frac{1}{1-\xid+\imath 0} +\imath \pi \delta(1-\xid) \right) 
+ 2 (1-\ep) \, (1-v_2)^{-1} (1-v_2 \, \xid) \bigg]~, \nn \\
\sigma_{\v,3}^{(1)} &=& - \sigma^{(0)} \, \gs^2 \, C_F \,  
\int d[\xi_{3,0}] \, d[v_3] \, \bigg[ 
 \frac{4\, \xi_{3,0}}{1+\xi_{3,0}} \left((1-v_3) (v_3^{-1}+\xi_{3,0})-\ep \right) \nn \\ 
&-& 2 (1-\ep) \, v_3^{-1} (1+(1-v_3) \, \xi_{3,0}) \bigg]~,
\label{eq:dualvirtualcrossTOTAL}
\eea
which fulfill 
\beq
\sigma_{\v}^{(1)} = \sum_{i=1,2} (\widetilde \sigma_{\v,i}^{(1)}+\bar \sigma_{\v,i}^{(1)}) \, + \sigma_{\v,3}^{(1)}.
\eeq
Finally, the UV cross-section is given by 
\beqn
\nn \sigma^{(1,\uv)}_{\v} &=& \sigma^{(0)} \, \gs^2 \,  C_F \, \int d[\xi_\uv] \, d[v_\uv] \, 
\bigg[ \frac{(1-\ep) \, (1-2v_\uv) \, \xi_\uv^3 \, ( 12 - 7 m_{\uv}^2 - 4 \xi_\uv^2 )} {(\xi_\uv^2 + m_{\uv}^2)^{5/2}} \\ 
&+& \frac{2\, \xi_\uv^2 ((1+2\ep)m_{\uv}^2 + 4\xi_\uv^2(1-\ep-3(2-\ep)v_\uv(1-v_\uv)))}{(\xi_\uv^2 + m_{\uv}^2)^{5/2}} \bigg]~.
\label{eq:sigmaUV}
\eeqn

Following the discussion presented in Section~\ref{sec:triangleREAL}, we also separate the real three-body phase-space 
to isolate the different collinear configurations; this motivates us to introduce
\beq
\widetilde \sigma_{\r,i}^{(1)} = \frac{1}{2s_{12}} \, \int d\Phi_{1\to 3} \,  |{\cal M}_{\qqg}^{(0)}|^2 \, \theta(y_{jr}' -y_{ir}' )~
\qquad i,j \in \{ 1,2\} \, , \quad i \neq j \ \ ,
\label{eq:realalqqg}
\eeq
that are analogous to \Eq{eq:separacionREALtoymodel} for the toy-model previously analysed 
and fulfil $\sum_i \widetilde \sigma_{\r,i}^{(1)} = \sigma_{\r}^{(1)}$. Explicitly,
\bea
\widetilde \sigma_{\r,1}^{(1)} &=& \sigma^{(0)} \, \gs^2 \,  C_F \, 
\int d[\xiu] d[v_1] \, {\cal R}_1(\xiu,v_1)  \, 
\frac{2 (1-\xiu)^{-2\ep}}{(1-(1-v_1)\, \xiu)^{-2\ep}}  \nn \\ &\times&
(v_1(1-v_1))^{-1} \,
\bigg[ \frac{(1-\xiu)^2+\left( (1-\xiu)^2\, (1-v_1)+v_1\right)^2}
{\xiu\, (1-(1-v_1)\, \xiu)^2}
-\ep \, \xiu \bigg]~, \nn \\ 
\widetilde \sigma_{\r,2}^{(1)} &=& \sigma^{(0)} \, \gs^2 \,  C_F \, 
\int d[\xid] d[v_2] \, \, {\cal R}_2(\xid,v_2) \, 
\frac{2 (1-\xid)^{-2\ep}}{(1-v_2\, \xid)^{1-2\ep}} \, 
\nn \\ &\times& (1-v_2)^{-1} 
\bigg[ \frac{(1-\xid)^2-\ep \left( (1-\xid)^2 \, v_2 + 1-v_2\right)^2}{(1-v_2\, \xid)^2} 
+\xid^2 \bigg]~,
\eea
which are obtained by applying the mappings defined in Eqs. (\ref{eq:mapping1})-(\ref{eq:EcuacionYprima2}) to~\Eq{eq:realtotal}, respectively. 
Since the total cross-section is given by
\beq
\sigma^{(1)} =  \sigma_{\v}^{(1,\r)} + \sigma_{\r}^{(1)}~,
\label{eq:totalcrosssecdecomposition}
\eeq
we put together $\widetilde \sigma_{\v,i}^{(1)}$ and $\widetilde \sigma_{\r,i}^{(1)}$ to define,
\beq
\widetilde \sigma_{i}^{(1)} = \widetilde \sigma_{\v,i}^{(1)} + \widetilde \sigma_{\r,i}^{(1)}~, 
\qquad i \in \{1,2\}~,
\label{eq:sigmatildeIdefinicion}
\eeq
that are finite in the limit $\ep \to 0$. Moreover, through the application of the momenta mapping and the separation of the integration region, there is a local cancellation of singularities which allows us to take the limit $\ep \to 0$ at integrand level, i.e. \emph{before integration}. 

From \Eq{eq:totalcrosssecdecomposition}, we appreciate that there are still some missing contributions to the cross-section. In fact, we must combine all the virtual terms and UV 
counter-terms that have not been included in the virtual dual cross-sections in~\Eq{eq:dualvirtualcross}. So, we define
\beq
\overline \sigma^{(1)}_{\v} = \left(\sum_{i=1,2} \, \bar \sigma_{\v,i}^{(1)} \right) \, +  \sigma_{\v,3}^{(1)} - \sigma_{\v}^{(1,\uv)}~,
\eeq
as the remnant virtual correction. It is worth appreciating that this expression admits a four-dimensional representation, 
which can be build applying the change of variables shown in Section~\ref{ssec:unify}. 
Finally, the calculation of the total integrals leads to 
\bea
\widetilde \sigma_{1}^{(1)} &=& \sigma^{(0)} \, \frac{\as}{4\pi} \,  C_F \, \left( 19 - 32 \, \ln{2} \right)~,  \nn \\
\widetilde \sigma_{2}^{(1)} &=& \sigma^{(0)} \, \frac{\as}{4\pi} \,  C_F \, \left( -\frac{11}{2} + 8 \, \ln{2} - \frac{\pi^2}{3}\right)~, \nn \\
\overline \sigma^{(1)}_{\v} &=& \sigma^{(0)} \, \frac{\as}{4\pi} \,  C_F \, \left( -\frac{21}{2} + 24 \, \ln{2} + \frac{\pi^2}{3}\right)~, 
\label{eq:resultadosfinales}
\eea
whose sum gives 
\beq
\widetilde \sigma_{1}^{(1)} + \widetilde \sigma_{2}^{(1)} +
\overline \sigma^{(1)}_{\v} = \sigma^{(0)} \, 3  \,  C_F \, \frac{\as}{4\pi}~, 
\eeq
that is in agreement with the $\ep \to 0$ limit of the result obtained through DREG. The crucial difference is that we get these results after the integration of four-dimensional expressions\footnote{Explicit four-dimensional expressions are shown
in Appendix~\ref{app:4d}.}, thus avoiding to deal with complicated $\ep$ expansions.

To conclude this section, let's notice that the procedure shown here could be simplified. 
In order to achieve an analytical computation it was necessary to split the integration region 
of the virtual corrections into different pieces. In a numerical implementation, this procedure 
will not be necessary. The loop integration will occur unrestricted, and the real corrections 
will be switch on to cancel the collinear and soft divergences in the region of the loop three-momentum 
where the respective momentum mapping conditions are fulfilled.

\section{Generalisation to multi-leg processes and NNLO}
\label{sec:global}

Assuming there are no initial-state partons (for instance in $e^+e^-$ annihilation) the generalisation to multi-leg 
processes is straightforward from the results presented in the previous sections. The implementation of the 
method in a Monte Carlo event generator is indeed simpler than presented before. As usual, the NLO 
cross-section is constructed from the one-loop virtual correction with $m$ partons in the final 
state and the exclusive real cross-section with $m+1$ partons in the final state
\beq
\sigma^\nlo = \int_{m} d\sigma_{\v}^{(1,\r)}+ \int_{m+1} d\sigma_{\r}^{(1)}~,
\eeq
where the virtual contribution is obtained from its dual representation 
\beq
d\sigma_{\v}^{(1,\r)} = \sum_{i=1}^N \int_\ell \,  2 \, {\rm Re} \, \la \M{0}_N|\M{1,\r}_N(\td{q_i}) \ra \, {\cal O}_{N}(\{p_j\})~.
\label{eq:nlov}
\eeq
In the above equation, $\M{0}_N$ is the $N$-leg scattering amplitude at LO, with $N>m$, and $\M{1,\r}_N$ is the renormalised 
one-loop scattering amplitude, which also contains the self-energy corrections of the external legs. 
On the other hand, ${\cal O}_{N}(\{{p}_j\})$ defines a given IR-safe physical observable by constraining the integration domain, 
for example, a jet function. 
The delta function $\td{q_i}$ symbolises the dual contribution with the internal momentum 
$q_i$ set on-shell. By \emph{renormalised}, we mean that appropriate UV counter-terms have been 
subtracted locally, according to the discussion presented in Section~\ref{sec:renorm}, including UV singularities of degree 
higher than logarithmic that integrate to zero, and the UV contributions from the wave function of the external 
particles, in such a way that only IR singularities arise in $d\sigma_{\v}^{(1,\r)}$. In~\Eq{eq:nlov}, we have also assumed a definite ordering of the external particles that leads to a definite
set of internal momenta $q_i$. This means, that the one-loop scattering amplitude $\M{1,\r}_N$ 
contains not only the contribution from the maximal one-loop $N$-point function, but also all the terms that 
can be constructed with the same set of internal momenta. Keeping this ordering is necessary to preserve 
the partial cancellation of singularities among dual contributions at integrand level. Obviously, all the possible permutations 
and symmetry factors have to be considered to obtain the physical cross-section. 

The real cross-section is given by 
\bea
\int_{m+1}\, d\sigma^{(1)}_{\r} &=& \sum_{i=1}^N \, \int_{m+1} \, |\M{0}_{N+1}(q_i,p_i)|^2 \, {\cal R}_i(q_i,p_i)\, {\cal O}_{N+1}(\{p'_j\})~, 
\label{eq:nlor}
\eea 
where the external momenta $\{p'_j\}$, the phase-space and the tree-level scattering amplitude $\M{0}_{N+1}$ are rewritten in 
terms of the loop three-momentum (equivalently, the internal loop on-shell momenta) and the external 
momenta $p_i$ of the Born process. The momentum mapping links uniquely the soft and collinear states of the real 
and virtual corrections. Therefore, if the physical observable is IR-safe then ${\cal O}_{N+1} \to {\cal O}_N$ in all the possible IR-degenerate configurations 
of the $(N+1)$-particle process. In this way, it is guaranteed that the simultaneous 
implementation of the real-emission terms with the corresponding dual contributions leads to an integrand level cancellation
of IR singularities.

\begin{figure}[t]
\begin{center}
\includegraphics[width=15cm]{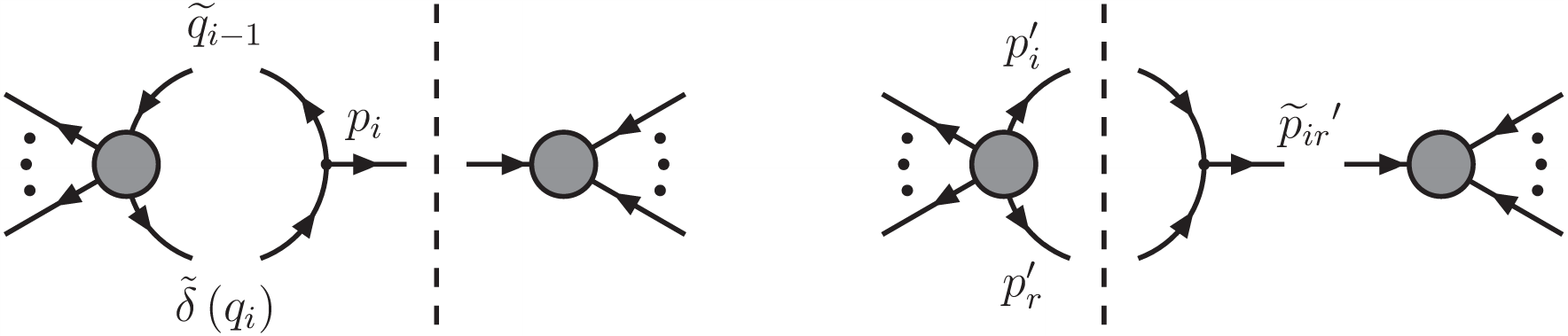}
\caption{\label{fig:collinear}
Factorisation of the dual one-loop and tree-level squared amplitudes 
in the collinear limit. The dashed line represents the momentum conservation
cut. Interference of the Born process with the one-loop scattering amplitude with 
internal momentum $q_i$ on-shell, $\M{1}_{N}(\td{q_i})\otimes \MD{0}_{N}$ (left), 
and interference of real processes with the parton splitting $p_{ir}'\to p_i'+p_r'$:
$\M{0}_{N+1}\otimes \MD{0}_{N+1, ir}$ (right). In this limit 
the momenta $q_{i-1}=q_i-p_i$ and $p_{ir}'$ become on-shell and the 
scattering amplitudes factorise.}
\end{center}
\end{figure}

Analogously to the dipole method~\cite{Catani:1996jh,Catani:1996vz}, in order to construct the momentum 
mapping between the $m$ and $m+1$ kinematics, we single out two partons for each contribution. 
The first parton is the {\it emitter} and the second parton is the {\it spectator}.
The difference with respect to the dipole formalism is that both the emitter and the spectator are initially related to external 
momenta of the virtual scattering amplitudes, and not to internal or external momenta of the real emission processes. 
Then, the loop three-momentum and the four-momenta of the emitter and the spectator
are used to reconstruct the kinematics of the corresponding real emission cross-section 
in the region of the real phase-space where the twin of the emitter decays to 
two partons in a soft or collinear configuration.  Explicitly, if the momentum of the final-state emitter 
is $p_i$, the internal momentum prior~\footnote{We assume that the internal and external 
momenta are ordered according to Fig.~\ref{fig:OneLoopScalar}.} 
to the emitter is $q_i$ and it is on-shell, and $p_j$ is the momentum of  the final-state 
spectator, then, the multi-leg momentum mapping is given by 
\bea
&& p_r'^\mu = q_i^\mu~, \nn \\
&& p_i'^\mu = p_i^\mu - q_i^\mu + \alpha_i \, p_j^\mu~, \qquad \alpha_i = \frac{(q_i-p_i)^2}{2 p_j\cdot(q_i-p_i)}~, \nn \\
&& p_j'^\mu = (1-\alpha_i) \, p_j^\mu~, \nn \\
&& p_k'^\mu = p_k^\mu~, \qquad \qquad \qquad \qquad k \ne i,j~.
\label{eq:multilegmapping}
\eea
The incoming initial-state momenta, $p_a$ and $p_b$, are left unchanged.  Momentum conservation is 
automatically fulfilled by~\Eq{eq:multilegmapping} because $p_i+ p_j +\sum_{k\ne i,j} p_k = p_i'+p_r'+p_j'+\sum_{k\ne i,j} p_k'$, 
and all the primed final-state momenta are massless and on-shell if the virtual unprimed momenta are also massless. 
The momentum mapping in~\Eq{eq:multilegmapping} is motivated by the general factorisation properties 
in QCD~\cite{Buchta:2014dfa,Catani:2011st}, and it is graphically explained in Fig.~\ref{fig:collinear}.

The emitter $p_i$ has the same flavour as $p_{ir}'$, the twin emitter or parent parton (called emitter in the dipole formalism) 
of the real splitting configuration that is mapped. The spectator $p_j$ is used to balance 
momentum conservation, and has the same flavour in the virtual and real contributions. 
As for the dipoles, there are alternatives to treat the recoiling momentum, but the option with a single spectator is the most suitable. 
The radiated particles, $p_i'$ and $p_r'$, might have a different flavour than the emitter $p_i$. If the emitter is a quark 
or antiquark, the role of $p_i'$ and $p_r'$ can be exchanged if the subindex $i$ is used to denote the flavour, 
as we did in~\Eq{eq:mapping2}. If the emitter is a gluon, the radiated partons are gluons or a quark-antiquark pair.

The momentum mapping in \Eq{eq:multilegmapping} is suitable in the region of the loop-momentum space 
where $q_i$ is soft or collinear with $p_i$, and therefore in the region of the real phase-space where 
$p_i'$ and $p_r'$ are produced collinear or one of them is soft. 
In~\Eq{eq:nlov}, we have introduced a complete partition of the real phase-space
\beq
\sum {\cal R}_i (q_i,p_i) = \sum \prod_{jk\ne ir} \theta(y_{jk}' -y_{ir}')   = 1~,
\eeq 
which is equivalent to divide the phase-space by the minimal two-body invariant $y_{ir}'$.
Since the real and virtual kinematics are related, the real phase-space partition defines 
equivalent regions in the loop three-momentum space. Notice, however, that we 
have not imposed these constrains in the definition of the virtual cross-section in~\Eq{eq:nlov}, 
as we did for the analytic applications presented in Sections~\ref{sec:triangleREAL} and~\ref{sec:gammatoqqbar}.
The actual implementation of the NLO cross-section in a Monte Carlo event generator is a single unconstrained 
integral in the loop three-momentum, and the phase-space with $m$ final-state particles. By virtue of 
the momentum mapping, real corrections are switch on in the region of the loop three-momentum where 
they map the corresponding soft and collinear divergences. This region is a compact region, 
and it is of the size of the representative hard scale of the scattering process. 
At large loop three-momentum only the virtual corrections contribute, and their UV 
singularities are subtracted locally by suitable counter-terms. These are the only 
counter-terms that are required for the implementation of the method, the IR singularities 
are unsubtracted because their cancellation is achieved simultaneously. 
The full calculation is implemented in four-dimensions, more precisely with the DREG 
parameter $\ep=0$. Moreover, there is no need to perform any tensor reduction 
in the calculation of the virtual contributions, hence avoiding the appearance of Gram determinants 
leading to spurious numerical instabilities. Integrable threshold singularities of the loop contributions
are also restricted to the physical compact region, and are treated numerically by contour 
deformation~\cite{Buchta:2015xda,Buchta:2015jea,Buchta:2015lva,Buchta:2015wna} in a Monte Carlo implementation. 
 
The case of lepton--hadron and hadron--hadron collisions deserves a comment apart. 
The cross-section is computed by convoluting the corresponding partonic cross-section 
with the process-independent parton distributions of the incoming hadrons.
Since the initial-state partons carry a well defined momentum, the partonic subprocesses 
are not collinear safe. By virtue of the universal factorisation properties of QCD
for massless incoming partons~\cite{Collins:1989gx}, 
the initial-state collinear singularities are factorised and reabsorbed in the definition of the 
non-perturbative parton distributions, and are removed from the partonic cross-section 
by suitable collinear counter-terms which are proportional to $1/\ep$ and the Altarelli-Parisi splitting 
functions~\cite{Sborlini:2013jba,Altarelli:1977zs,deFlorian:2015ujt,Sborlini:2014kla,Sborlini:2014mpa,Catani:2003vu}.
The initial-state collinear counter-terms are convoluted with the Born cross-section. However, 
their standard form is not suitable in the LTD approach because that convolution 
is a single integral in longitudinal momentum with Born kinematics. Its unintegrated form,
which should depend also on the transverse momentum of the real radiation, is necessary, 
and will be presented in a future publication.  
 
The extension to NNLO, although not straightforward, can instinctively be anticipated and will be 
treated also in a future publication. The NNLO cross-section consists of three contributions 
\beq
\sigma^{\rm NNLO} = \int_{m} d\sigma_{\v\v}^{(2)} + \int_{m+1} d\sigma_{\v\r}^{(2)} + \int_{m+2} d\sigma_{\r\r}^{(2)}~,
\eeq
where the double virtual cross-section $d\sigma_{\v\v}^{(2)}$ receives contributions from 
the interference of the two-loop with the Born scattering amplitudes, and the square 
of the one-loop scattering amplitude with $m$ final-state particles, the virtual-real cross-section 
$d\sigma_{\v\r}^{(2)}$ includes the contributions from the interference of one-loop and tree-level scattering 
amplitudes with one extra external particle, and the double real cross-section $d\sigma_{\r\r}^{(2)}$ are 
tree-level contributions with emission of two extra particles. 
The LTD representation of the two-loop scattering amplitude is obtained by setting two 
internal lines on-shell~\cite{Bierenbaum:2010cy}. It leads to the two-loop dual components
$\la \M{0}_N|\M{2}_N(\td{q_i},\td{q_j})\ra$, while the two-loop momenta of the squared one-loop 
amplitude are independent and generate dual contributions of the type 
$\la \M{1}_N(\td{q_i})|\M{1}_N(\td{q_j})\ra$. In both cases, we have at our disposal two independent
loop three-momenta and $m$ final-state momenta, from where we can reconstruct the 
kinematics of the one-loop corrections entering $d\sigma_{\v\r}^{(2)}$, and the 
tree-level corrections in $d\sigma_{\r\r}^{(2)}$.

\section{Conclusions and outlook}
\label{sec:conclusions}

In this article, we carefully discussed the implementation of a novel algorithm to compute higher-order 
corrections to physical observables. This method is based on the LTD theorem, which states that virtual 
contributions can be expressed as the sum over single-cut (at one-loop) or dual integrals, whose structure closely 
resembles real-emission amplitudes. We exploit this knowledge to perform an integrand-level combination
of real and virtual terms, which leads to a fully local cancellation of singularities and allows to implement 
the calculation without making use of DREG.

One of the interesting uses of LTD lies in the possibility of exploring the causal structure of virtual contributions
and disentangling their singularities. In particular, we applied this technique to prove that the IR divergences are 
generated in a compact region of the loop-momentum space. This is a crucial fact in order to achieve the 
real--virtual cancellation of singularities in IR-safe observables. 

To illustrate the importance of the compactness of the IR singular regions, we studied the NLO corrections to 
a $1 \to 2$ process in the context of a toy scalar model. 
By using suitable momentum mappings, we generated $1 \to 3$ on-shell massless 
kinematics from the $1 \to 2$ process and the loop three-momentum. 
These mappings relate exactly the integration regions where the singularities are originated. 
In this example, we distinguished two regions in the real-emission phase-space and defined suitable
mappings to cover them in the dual space of the loop three-momentum.
In this way, the combination of dual and real contributions led to expressions that are 
integrable in four-dimensions;  the remainders of the virtual part were also represented 
by a four-dimensional integral. The algorithm is called unsubtracted because the summation 
over degenerate soft and collinear final-states is performed thanks to these momentum 
mappings, then making unnecessary the introduction of IR subtractions. 

On the other hand, we investigated the cancellation of UV divergences at integrand level. We started with the 
simplest example of a massless scalar two-point function.  Using the ideas presented in 
Ref.~\cite{Becker:2010ng}, we obtained the dual representation of the local UV counter-term that 
exactly cancels the divergences in the high-energy region of the loop momentum, 
rendering an integrable representation in four-dimensions. We also extended the procedure to deal 
with arbitrary scattering amplitudes, and provided a subtle physical interpretation of the 
energy scale entering the UV counter-term as renormalisation scale. 

Then, we applied the IR unsubtraction LTD algorithm to the physical process $\gamma^* \to q \bar{q}(g)$ 
at NLO in QCD. In the first place, we found the dual representation of the virtual contribution and made 
use of the momentum mappings to perform the real--virtual combination. It is worth appreciating that
we had to include also self-energy corrections to the external on-shell legs,
while they are usually ignored because their integrated form vanishes in DREG due to the lack 
of physical scales. Since our approach explicitly splits the IR and UV regions in the dual integrals, 
it was necessary to disentangle their IR/UV behaviour.  In this way, we computed the full NLO 
correction to $\gamma^* \to q \bar{q}(g)$ by making use of purely four-dimensional expressions
also in a physical application.

The generalisation of the algorithm to deal with multi-particle processes is quite straightforward, at least 
when only final-state singularities take place. Essentially, the real emission phase-space must be split to 
isolate the different collinear configurations and, then, a proper momentum mapping for each of these 
configurations is defined.  Combining the dual integrands with the real matrix elements in the corresponding 
regions leads to integrable expressions in four-dimensions. 
The cancellation of UV divergences is done following the same ideas as in the $1 \to 2$ case. 
To conclude, we succinctly sketched the extension of the algorithm to cure initial-state collinear singularities, and 
discussed briefly its generalisation to NNLO.

In summary, in this work we presented explicitly a well-defined algorithm that allows to override DREG by 
exploiting LTD. It constitutes a new paradigm in perturbative calculations because it takes 
advantage of combining directly real and virtual corrections in an integrable four-dimensional 
representation, while providing an easy physical interpretation of the singularities of the scattering amplitudes
and unveiling their hidden nature.

\section*{Acknowledgements}
 
This work has been supported by CONICET Argentina,
by the Spanish Government and ERDF funds from the European Commission 
(Grants No. FPA2014-53631-C2-1-P and SEV-2014-0398)
and by Generalitat Valenciana under Grant No. PROMETEOII/2013/007. FDM acknowledges suport from Generalitat
Valenciana (GRISOLIA/2015/035). The work of RJHP is partially supported by CONACyT, M\'exico.

\appendix
\section{The dual integration measure}
\label{app:measure}

Starting from the expressions for the loop integration measure and the on-shell delta function given in Section~\ref{sec:notationINTRO}, we define 
\beq
\widetilde{c}_\Gamma = \frac{c_\Gamma}{\cos(\pi \epsilon)}~, \qquad
c_\Gamma = \frac{\Gamma(1+\epsilon)\Gamma^2(1-\epsilon)}
{(4\pi)^{2-\epsilon}\Gamma(1-2\epsilon)}~, 
\label{eq:cgamma}
\eeq
as the phase-space and one-loop volume factors, respectively. Notice that 
\beq
\widetilde{c}_\Gamma \equiv c_\Gamma\, + \, {\cal O}(\ep^2)~, \qquad
c_\Gamma \equiv \frac{(4\pi)^{\epsilon-2}}{\Gamma(1-\epsilon)} \, + \, {\cal O}(\ep^3) ~, 
\label{eq:cgammaexpanded}
\eeq
which implies that at NLO, we can interchange these prefactors without altering the final expressions up to ${\cal O}(\epsilon)$. 

On the other hand, using spherical coordinates in $d$-dimensions, the dual integration measure is rewritten as
\bea
\int_{\ell} \, \td{q_i} &=& \frac{\mu^{2\ep}}{(2\pi)^{d-1}}
\int \, d^d q_i \, \theta(q_{i,0}) \, \delta(q_i^2-m_i^2) = 
\frac{\mu^{2\ep}}{(2\pi)^{d-1}} \int \, \frac{(\qb_i^2)^{1-\ep}}{2\, q_{i,0}^{(+)}}   
\, d|\qb_i| \, d\Omega_i^{(d-2)}~,
\eea
where $q_{i,0}^{(+)} = \sqrt{\qb^2_i+m_i^2-\imath 0}$. If the azimuthal integration is trivial, 
with $\cos \theta_i = 1-2 v_i$ the cosinus of the polar angle, 
the solid angle is given by 
\beq
d\Omega_i^{(d-2)}=\frac{(4\pi)^{1-\epsilon}}{\Gamma(1-\epsilon)}
\int_0^1 \, d[v_i]~, \qquad
d[v_i]  = (v_i(1-v_i))^{-\epsilon} \, dv_i~,
\eeq
that leads to
\beq
\int_{\ell} \, \td{q_i}  = 
\frac{\mu^{2\ep}(4\pi)^{\ep-2}}{\Gamma(1-\ep)} \int \, \frac{(4 \qb_i^2)^{1-\ep}}{q_{i,0}^{(+)}}   
d|\qb_i| \, d[v_i]~. 
\eeq
Since in this work we deal with massless internal propagators, then $q_{i,0}^{(+)} = \sqrt{\qb_i^2-\imath 0}$ and the dual 
integration measure simplifies to 
\beq
\int_{\ell} \, \td{q_i}  = 
\frac{\mu^{2\ep}(4\pi)^{\ep-2}}{\Gamma(1-\ep)} \int \, (2 q_{i,0})^{1-2\ep}  \, d(2 q_{i,0}) \, d[v_i]~,  
\eeq
which justifies the notation established in \Eq{eq:measureDREG}.

\section{Phase-space}
\label{app:phasespace}

Following the notation introduced in~\Eq{eq:tddefinition}, 
the $d$-dimensional phase-space associated to a $N$-leg scattering process  
with $m$ final-state particles is given by the expression
\beqn
d\Phi_{m} &=& \mu^{d-4} 
\left( \prod_{i=1}^m \, \int_{p_i}\, \td{p_i} \right)\, (2\pi)^d \, \delta^{(d)}\left(\sum_{i=1}^m p_i - p_a\right)~,
\label{GenericPHASESPACE}
\eeqn
with $p_a$ the sum of the incoming momenta for either $1\to m$ or $2\to m$ processes. 
The total volume of the $1 \to 2$ phase-space is given by the well-known formula 
\beqn
\int \, d\Phi_{1\to 2} &=& \frac{\Gamma(1-\ep)}{2(4\pi)^{1-\ep} \,\Gamma(2-2\ep)}
 \left(\frac{s_{12}}{\mu^2}\right)^{-\epsilon}~, 
\label{Medida12INTEGRADA}
\eeqn
where $s_{12}=p_{12}^2$ is the invariant mass of the decaying particle. For the NLO real-radiation correction to $1 \to 2$ 
process we need to deal with $1 \to 3$ decays; the corresponding phase-space is
\beqn
d\Phi_{1\to 3} &=& \frac{s_{12}}{2 \, (4\pi)^{3-2\epsilon} \Gamma(2-2\ep)} \, \left(\frac{s_{12}}{\mu^2}\right)^{-2\epsilon}\, 
(y_{12}' \, y_{1r}' \, y_{2r}')^{-\epsilon} \, dy_{1r}' \, dy_{2r}'~,
\label{Medida13ORIGINAL}
\eeqn
where the parametrisation used is the one introduced in Eqs. (\ref{eq:EcuacionYprima1}) and (\ref{eq:EcuacionYprima2}). 
Because we are interested in the relative factors of $d\Phi_{1\to 3}$ compared with $d\Phi_{1\to 2}$, \Eq{Medida13ORIGINAL} 
can be rewritten as
\beqn
d\Phi_{1\to 3} &=& \frac{(4\pi)^{\ep-2} \, s_{12}}{\Gamma(1-\ep)}\, 
\left(\frac{s_{12}}{\mu^2}\right)^{-\epsilon}\, \left(\int \, d\Phi_{1\to 2} \right) \, 
(y_{12}' \, y_{1r}' \, y_{2r}')^{-\epsilon}\, dy_{1r}' \, dy_{2r}'~.
\label{Medida13ORIGINALrelativa}
\eeqn
Notice that the kinematical configuration in the TL-region is given by $p_3 \to p_1 + p_2$ (or $p_3 \to p_1' + p_2' + p_r'$ for the real term), with $p_3$ a massless off-shell particle whose virtuality is $p_3^2=s_{12}>0$. The integrand is considered Lorentz invariant, so we integrated out angular degrees of freedom associated with the extra-dimensions.

Finally, let's recall another useful technique for phase-space integration and $\ep$-expansion of the results, which relies in the application of distributions~\cite{Frixione:1995ms} to rewrite the DREG measure. The basic formula is given by
\beqn
\nn \int_0^1 dx \, x^{-1+a\ep} \, f(x,\epsilon) &=& \int_0^1 dx \, \left[\frac{(x_C)^{a \,\ep}}{a \ep} \delta(x) + \left(\frac{1}{x}\right)_C + a \epsilon \left(\frac{\ln{x}}{x}\right)_C\right] f(x,\epsilon) + {\cal O}(\ep^2), \\
\label{Distro1}
\eeqn
where $x_C \in (0,1]$ is an arbitrary cut and $f$ is a generic test function without singularities at $x=0$ and without $\ep$-poles. The $C$-distributions are defined according to
\beqn
\int_0^1 dx \, \left(\frac{\log^n(x)}{x}\right)_C \, f(x,\epsilon) &=& \int_0^1 dx \, \log^n(x) \, \frac{f(x,\epsilon)-f(0,\epsilon)\, \theta(x_C-x)}{x} \, ,
\label{Distro2}
\eeqn
i.e. the cancellation of the integrand is forced in a neighbourhood of the singular point $x=0$. Notice that the test function must be an entire function of $\ep$ in order to avoid additional $\ep$-poles. Besides that, Eqs. (\ref{Distro1}) and (\ref{Distro2}) can be adapted to several domains and measures. In particular, the $\ep$-expansion of the phase-space at large loop momentum becomes expressible as
\beqn
\nn \int_\ell \td{q_i} f(q_i,\ep) &\equiv& 4 c_\Gamma \, \int q_{i,0} \, dq_{i,0} \, \int dv \, \bigg\{
\left[ 1 + 2 \log(q_{i,0}) 
\left(v \, \delta(v) + (1-v)\, \delta(1-v) \right)  \right] f(q_i,0)
\\ &-& 
\left(v \, \delta(v) + (1-v)\, \delta(1-v) \right) \, \left.\frac{\partial f}{\partial \epsilon}\right|_{\ep=0} \bigg\} + {\cal O}(\ep) \, ,
\label{ExpansionMEDIDALOOP} 
\eeqn
under the assumption that $f(q_i, \ep)$ is an entire function of $\ep$ and it has a vanishing soft limit (i.e. $f\to 0$ as $q_{i,0} \to 0$). Appreciate the presence of extra-contributions given by the collinear residues at $v=0$ or $v=1$, as well as some additional terms introduced by the linear $\ep$ dependence of the integrand.

\section{NLO corrections to $\gamma^* \to q \bar q (g)$: explicit 4D formulae}
\label{app:4d}

In this Appendix, we collect the four-dimensional representation of the integrands associated to the integrals 
in~\Eq{eq:resultadosfinales}. Explicitly, we have
\beqn
\nn \widetilde \sigma_{1}^{(1)} &=& \sigma^{(0)} \, \frac{\as}{4\pi} \,  C_F \, \int_0^1 d\xi_{1,0} \, \int_0^{1/2} dv_1 \, 4\, {\cal R}_1(\xi_{1,0},v_1) \, 
\left[ 2 \left( \xiu - (1-v_1)^{-1} \right) - \frac{\xiu (1-\xiu)}{\left(1-(1-v_1) \, \xiu \right)^2} \right]~,  
\label{eq:APPENDIXexplicitsigma1}
\\ \nn \widetilde \sigma_{2}^{(1)} &=& \sigma^{(0)} \, \frac{\as}{4\pi} \,  C_F \, \int_0^1 d\xi_{2,0} \, \int_0^1 dv_2 \, 2 \, {\cal R}_2(\xi_{2,0},v_2) \, 
(1-v_2)^{-1}\,  \Bigg[ \frac{2 \, v_2 \, \xid \left(\xid (1-v_2) - 1\right)}{1-\xid}  \\ &&
- 1 + v_2\, \xid  + \frac{1}{1-v_2 \, \xid} \left(\frac{(1-\xid)^2}{(1-v_2 \, \xid)^2} + \xid^2\right)  \Bigg]~,
\label{eq:APPENDIXexplicitsigma2}
\eeqn
for the real-virtual combinations, and
\beqn
\nn \overline \sigma^{(1)}_{\v} &=& \sigma^{(0)} \, \frac{\as}{4\pi} \,  C_F \, \int_0^{\infty} d\xi \, \int_0^1 dv \, 
\Bigg\{ - 2 \, \left(1-{\cal R}_1(\xi,v)\right) \, v^{-1}(1-v)^{-1} \, \frac{\xi^2 (1-2 v)^2+1}{\sqrt{(1+\xi)^2 - 4 v\, \xi}} 
\\ \nn &+& 2 \, \left(1-{\cal R}_2(\xi,v)\right) \, (1-v)^{-1}\,  \left[2 \, v \, \xi \, \left(\xi (1-v) - 1\right)
\left(\frac{1}{1-\xi+\imath 0} + \imath \pi \delta(1-\xi)  \right) - 1 + v \, \xi \right] 
\\ \nn &+& 2\, v^{-1} \left( \frac{\xi (1-v)(\xi (1-2v)-1)}{1+\xi} +1\right) 
- \frac{(1-2v) \, \xi^3 \, ( 12 - 7 m_{\uv}^2 - 4 \xi^2 )} {(\xi^2 + m_{\uv}^2)^{5/2}} \\ 
&-& \frac{2\, \xi^2 (m_{\uv}^2 + 4\xi^2(1-6v(1-v)))}{(\xi^2 + m_{\uv}^2)^{5/2}} \Bigg\}~, 
\label{eq:APPENDIXexplicitremnant}
\eeqn
the dual virtual remnant. In the previous expression, we have identified all the integration variables, 
$\xi=\xid=\xit=\xi_{\uv}$ and $v=v_2=v_3=v_{\uv}$, while $(\xiu,v_1)$ are expressed in terms of 
$(\xit,v_3)$  by using the change of variables in \Eq{eq:changexi10}. 
The integration regions are defined through~\Eq{eq:theta}, and we use~\Eq{eq:errequeerre}
to simplify the analytic integration. Notice that the integrand of the dual virtual remnant behaves as 
\beq 
\frac{d \, \overline \sigma^{(1)}_{\v}}{d\xi\, dv} \propto \frac{1-2v}{\xi^2} + {\cal O}(\xi^{-3})~,
\eeq
in the high-energy limit, and as ${\cal O}(\xi^{-3})$ after angular integration,
thanks to the UV counter-term that subtracts up to linear UV divergences locally. 




\end{document}